\documentclass[a4paper,article,10pt]{memoir}
\pdfoutput=1
\usepackage[russian,english]{babel}
\usepackage{CCLAuthor}
\usepackage[utf8x]{inputenc}
\usepackage[T2A]{fontenc}
\usepackage[reqno]{amsmath}% AMS-LaTeX-first
\usepackage{amssymb}
\usepackage{amsthm}
\theoremstyle{plain}

\theoremstyle{definition}

\numberwithin{theorem}{chapter}% AMS-LaTeX-last
\usepackage[round,authoryear]{natbib} % bibliography
\usepackage[]{graphicx}               % graphics
\newcommand{\crowd}{u}
\newcommand{\rr}[1]{{\normalfont\textrm{#1}}}
\graphicspath{ {./figures/} }
\begin{document}
%
% ----------------------------------------------
%
% declare the TITLE of your contribution
%
\title{Pedestrians moving in dark: Balancing measures and playing games on lattices}
%
% AUTHOR(s) and AFFILIATION(s)
%
\author{%
  Adrian Muntean\textsuperscript{*\dag}
  and
  Emilio N. M.\ Cirillo\textsuperscript{\ddag}
  \\
  Oleh Krehel\textsuperscript{*}
  and
  Michael B\"ohm\textsuperscript{**}
  \\  \smallskip\small% some space
  \textsuperscript{*} % first author
  CASA - Center for Analysis, Scientific computing and Applications,
  Department of Mathematics and Computer Science,
  Eindhoven University of Technology, The Netherlands
  \\
  \textsuperscript{\dag} % second author
  Institute for Complex Molecular Systems (ICMS), Eindhoven University of Technology, The Netherlands
  \\
  \textsuperscript{\ddag} % a common affiliation
  Dipartimento di Scienze di Base e Applicate per l'Ingegneria, Sapienza Universit\`a di Roma, Italy
  \\
  \textsuperscript{**}
  Zentrum f\"ur Technomathematik, Fachbereich Mathematik und Informatik, Universit\"at Bremen, Germany
}
%
% typeset the title
%
\maketitle
%
% ----------------------------------------------
%
% ABSTRACT: comment lines between ABS-first and ABS-last
% if your contribution doesn't include an abstract
%
% ABS-first
% \vskip-2\baselineskip
\noindent \begin{abstract}
  We present two conceptually new modeling approaches aimed at
  describing the motion of pedestrians in obscured corridors: \begin{itemize}
  \item[(i)] a
    Becker-D\"{o}ring-type dynamics and
  \item[(ii)] a probabilistic cellular automaton model.
  \end{itemize}
  In both models the group formation is affected by a threshold. The pedestrians are
  supposed to have very limited knowledge about their current position and their
  neighborhood; they can form groups up to a certain size and they can leave
  them. Their main goal is to find the exit of the corridor.

  Although being of mathematically different character, the discussion of both
  models shows that it seems to be a disadvantage for the individual to adhere
  to larger groups.

  We illustrate this effect numerically by solving both model systems. Finally
  we list some of our main open questions and conjectures.

  % We present two conceptually new modeling strategies able to describe
  % the motion of pedestrians in obscured corridors: (1) a
  % Becker-D\"oring-type dynamics for mass measures and (2) a lattice
  % model with threshold dynamics. The dynamics of the two models is
  % affected by the presence of a threshold, modeling group
  % formation. Although essentially different as mathematical structure,
  % both models underline the same idea: From the perspective of the total
  % evacuation flux, if large pedestrian groups move in dark corridors
  % with {\em a priori} unknown location of the exists and pedestrians can
  % not communicate with each other, then it is better to not behave
  % gregariously (to not adhere to large groups). We illustrate this
  % effect numerically by solving both the lattice automaton as well as
  % the Becker-D\"oring system. Finally, we list our main open questions
  % and conjecture on the existence of a two-scale measure-valued
  % formulation coupling the two modeling strategies into an unified
  % approach.
\end{abstract}
% ABS-last

\CCLsection{Introduction}\label{introduction}

Social mechanics is a topic that has attracted the attention of
researchers for more than one hundred years; see e.g. \citep{Haret,Apuntes}.
% There is a tacit belief that one can aim  at getting a deterministic answer to questions
% referring to the stochastic transport and local dynamics of pedestrians,
% especially in the high density regime, where the stochasticity in the system
% can be  in principle perceived as a controllable perturbation.
% Some crowd dynamics scenarios fit to this picture, some other do not.
% Driven by practical concerns, we decide to study  evacuation scenarios,
% but some of these ideas can be applied to other logistic situations
% when human-human or human-structure interactions play a role.
A large variety of existing models are able to describe the dynamics
of pedestrians driven by a {\em desired velocity} towards clearly defined exits.
But how can we possibly describe the motion of pedestrians
when the exits are not clearly defined, or even worse, {\em what if the exits are not visible}?

This paper is inspired by a practical evacuation scenario.
Some of the existing models are geared towards describing
the dynamics of pedestrians with somehow given, prescribed or,
at least, desired velocities or spatial fluxes towards an exit the location of which is,
more or less, known to the pedestrians\footnote{Efficient evacuation
of humans from high--risk zones is a very important issue
cf. \cite{Armin}. The topic is very well studied by large communities
of scientists ranging from logistics and transportation, civil and
fire engineering, to theoretical physics and applied mathematics.
Models (deterministic or stochastic) succeed to capture basic
behaviors of humans walking within given geometries towards {\em a
priori} prescribed exits. Typical classes of crowd dynamics models
include social force/social velocity models (cf. e.g.
\cite{HelbingMolnar}, \cite{PiccoliTosinMeasTh}, \cite{EversMuntean}),
simple asymmetric exclusion models (see chapters 3 and 4 from
\cite{Schadschneider2011} as well as references cited therein),
cellular automaton-type models \cite{Kirchner,Guo}, etc.; a detailed
classification of pedestrian models, see \cite{Schadschneider2011},
e.g. }.  We focus on modeling basic features which we assume to be
influencing the motion of pedestrians in regions with reduced or no
visibility\footnote{In recent years, high-rise buildings claim
steadily increasing numbers of victims in evacuations. Most victims
were due to the reduced visibility by fire smoke; see
\cite{Jin78,Jin85}.  In the future, most likely one will insist also
on building underground, so the potential of smoke victims further
increases. We refer the reader to \cite{Bauke} for a recent literature
review.}.  Our scenario is the following: A large number of
pedestrians, generally denoted by $Y$, is supposed to move through an
obscured corridor,$\Omega$. Due to the lack of visibility (e.g.
smoke, fog, darkness, etc.\footnote{Think about an evacuation in a
metro in which there is smoke and/or no light, etc.}) the $Y$'s cannot
see the exit. We allow for some sort of "buddying": If $Y$'s hit each
other they might decide to form a group. For practical reasons, we
limit the size of such groups by a threshold $T$. As transport
mechanism, we assume a very mild diffusion-like motion which is not
connected with the location of the exit. To model this situation, we
take two different routes by introducing and discussing:
\begin{itemize}
\item[(1)] a Becker-D\"oring-type system of balance equations for mass measures (see Appendix \ref{BB} for a derivation)
\item[(2)] a lattice model for an interacting particle system with threshold dynamics.
\end{itemize}
The two approaches are conceptually different. They consider from two
different perspectives the concept of {\em group} (social
collectivity). In the following sections, we approximate the
corresponding dynamics for evacuation scenarios similar to those
described in \cite{Jun} and \cite{Zheng}, for instance. In the first
approach, the {\em group feature} is imbedded in a size-dependent mass
measure and the evolution will be dictated by the conservation
equation of the respective measure (balancing the size-dependent
density).  In the second approach, we use a threshold to allow finite
non-exclusion per site in a lattice automaton for the self-propelled
particles (i.e. the pedestrians). We suspect however that connections
between (1) and (2) might exist, but we don't expect that the
mean-field limit of (2) is (1) (cf. e.g. \cite{Errico}).

Whatever route we take, our  central questions are:
\begin{center}
  \begin{itemize}
  \item[(Q1)] {\em How do pedestrians choose their path and speed when they are about  to move through regions with no visibility? }
  \item[(Q2)] {\em Is group formation (e.g. buddying) the right strategy to move through such uncomfortable zones  able to ensure exiting
      within a reasonable time?}
  \end{itemize}
\end{center}
Answers to (Q1) and (Q2) are largely unknown.
Group psychology (compare e.g.  \cite{LeBon,Curseu} and \cite{DyerJohanssonHelbing})
lacks extensive  experimental observations, and, due to absence of meaningful statistics,
nothing can be really concluded. The "groups" we study here are expected
to be highly unstable and therefore they only remotely resemble
the well-studied swarming patterns typically observed in nature
by fish and or birds communities
(see e.g. the 4--groups taxonomy in
\cite{Topaz}, namely swarm, torus, dynamic parallel groups, and highly parallel groups).
% On top of this, note that concepts like leadership (see
% e.g. \cite{CouzinNature}) or motion fluidization by favorizing lanes
% formation (see e.g. \cite{HelbingVicsek}) are simply not applicable in
% this context.

The basic idea is the following:
In the situation we are modeling, neighbors (both individuals or groups)
can not be visually identified by the individuals in motion,
so that basic mechanisms like attraction to a group, tendency to align,
or social repulsion are negligible and individuals have to live with
``preferences".

The paper is structured as follows: We start off with a continuum
model describing the mesoscopic dynamics of groups in Section
\ref{BDa}. After giving the set of governing equations in Section
\ref{gv}, we illustrate numerically the observed threshold effects at
such mesoscopic level in Section \ref{thr}.  Appendix \ref{BB}
contains a formal derivation in terms of mass measures of the
Becker-D\"oring-like system proposed here.  As next step, we propose a
lattice model to capture the microscopic dynamics, see Section \ref
{s:random}. The model detailed in Section \ref{mosca} is illustrated
numerically in Section \ref{play}. We conclude by enumerating a set of
basic questions that are for the moment open (see Section
\ref{s:open}) on the behavior of both interacting particle systems and
structured densities with threshold effects.

\CCLsection{Becker-D\"oring grouping in action}\label{BDa}
\CCLsubsection{From interacting colloids to group dynamics}\label{gv}

Inspired by the modeling of charged colloids transport in porous media
(see e.g. \cite{KrehelKnabner,Ray}), we consider now a system of
reaction-diffusion equations describing the aggregation and
dissolution of groups; the $i$th variable in the vector of unknowns
represents the specific size of the subgroup $i$ (density of the
$i$-mer $\crowd_i$). Here $\crowd_1$ -- density of crowds of group
size one (individuals), $\crowd_2$ -- density of groups of size two,
and so on until $\crowd_N$ are the corresponding Radon-Nikodym
derivatives of suitable measures
%$\mu_Y(t,x,\{1\})$, $\mu_Y(t,x,\{2\})$, $\mu_Y(t,x,\{3\})$, ..., $\mu_Y(t,x,\{T\})$ with respect to the space-time Lebesgue measure $\lambda_{tx}$
(see Appendix \ref{BB} for details).
For convenience, we take here $T:= N$, the biggest group size.

The following equations describe our system:
\begin{eqnarray}
  &\partial _t\crowd_1+\nabla \cdot (-d_1\nabla \crowd_1)=-\crowd_1\sum_{i=1}^{N-1}\beta_i\crowd_i
  +\sum_{i=2}^N\alpha_i\crowd_i-\beta_1\crowd_1\crowd_1+\alpha_2 u_2 \label{eq1} \\
  &\partial _t\crowd_2+\nabla \cdot (-d_2\nabla \crowd_2)=\beta_1\crowd_1\crowd_1-\beta_2\crowd_2\crowd_1+\alpha_3\crowd_3-\alpha_2\crowd_2\\
  &\vdots\\
  &\partial _t\crowd_{N-1}+\nabla \cdot (-d_{N-1}\nabla \crowd_{N-1})=\beta_{N-2}\crowd_{N-2}\crowd_1-\\
  &-\beta_{N-1}\crowd_{N-1}\crowd_1
  +\alpha_{N}\crowd_{N}-\alpha_{N-1}\crowd_{N-1}\\
  &\partial _t\crowd_N+\nabla \cdot (-d_N\nabla \crowd_N)=\beta_{N-1}\crowd_{N-1}\crowd_1-\alpha_N\crowd_N. \label{eq6}
\end{eqnarray}
This system of partial differential equations indicates that groups
diffuse inside $\Omega$. If the groups meet each other, then they
start to interact via the mechanism suggested by the right-hand side
of the system (aggregation or degradation being the only allowed
interaction behaviors).
We take as boundary conditions
\begin{eqnarray}
  &u_1=0\qquad\text{on }\Gamma_D\\
  &-d_1\nabla u_1\cdot n=0\qquad\text{on }\partial\Omega\setminus\Gamma_D\\
  &-d_i\nabla u_i\cdot n=0\qquad\text{on }\partial\Omega,i\in\{2,\ldots,N\},
\end{eqnarray}
while  the initial conditions at $t=0$ are
\begin{eqnarray}
  &u_1=M\qquad\text{in }\Omega\label{mem}\label{iniM}\\
  &u_i=0\qquad\text{ in }\Omega,i\in\{2,\ldots,N\}.
\end{eqnarray}
These boundary conditions model the following scenario: Only the
population of size one are allowed to exit, all the other groups need
to split in smaller groups close to $\Gamma_D$.  In (\ref{iniM}),
$M>0$ denotes the initial density of individuals, the total mass [of
pedestrians] in the system being $\int_\Omega \sum_{i=1}^N iu_i$. The
total mass at $t=0$ is $M|\Omega|$.  Note that (\ref{mem}) indicates
that, initially, groups are not yet formed. Group formation happens
here immediately after the initial time.  As transport mechanism, we
have chosen to use Fickian diffusion fluxes to model the mesoscopic
erratic motion of the crowd [with all its $N$ group structures] inside
the corridor $\Omega$.
% The source term $s$ represents the exiting mass that is once again redistributed among $u_1$
% uniformly throughout $\Omega$ (so that mass is conserved and we can observe some steady-state behavior).

Similarly to the case of moving colloidal particles in porous media
(cf. for instance \cite{KrehelKnabner} and references cited therein),
we take as reference diffusion coefficients the ones given the
Stokes-Einstein relation, i.e.  the diffusion coefficient of the
social conglomeration is inversely proportional to its size as
described by $d_i:=\frac{1}{\sqrt[3]{i}}$ (which would correspond to
the colloidal particles diffusion in a 3D confinement) for any $i\in
\{1,\dots, N\}$; see for instance \cite{Stokes-Einstein}. In contrast
to the case of transport in porous media, we assume that no
heterogeneities are present inside $\Omega$.  Consequently, the
diffusion coefficients are taken here to be independent of the space
and time variables.  If heterogeneities were present (like it is
nearly always the case e.g. in shopping malls), then one needs to
introduce concepts like local porosity and porosity measures as in
\cite{EversMuntean}; see \cite{Peruani} for a related scenario
discussing stochastically interacting self propelled particles within
a heterogeneous media with dynamic obstacles. We restrict ourselves
here to the case of homogeneous corridors.

We take the degradation (dissociation, group splitting) coefficients
$\alpha_i>0$ ($i\in\{2,\dots, N\}$) as being given constants, while
for the aggregation coefficients we use the concept of {\em social threshold}.
We define
\begin{equation}\label{defT}
  \beta_i:=
  \begin{cases}
    i&i<T\\
    1&\text{otherwise,}
  \end{cases}
\end{equation}
where $T\in (0,\infty)$ is the social threshold. Essentially, using
(\ref{defT}) we expect that the choice of $T$ essentially limits the
size of groups that can be formed by means of this
Becker-D\"oring-like model. In other words, even if large values of
$N$ are allowed (say mimicking $N\to \infty$) most likely groups of
sizes around $\left \lfloor T \right \rfloor$ will be created; here
$\left \lfloor p \right \rfloor$ denotes the integer part of
$p\in\mathbb{R}$.

\CCLsubsection{Threshold effects on mesoscopic group formation}\label{thr}

For the numerical examples illustrated here,
we consider $N=20$ species waking inside the corridor  $\Omega=(0,1)\times(0,1)$.
On the boundary $\partial\Omega$, we design the door $\Gamma_D=\{(x,y):\:x=0,y\in [0.4,0.6]\}$,
while the rest of the boundary $\partial\Omega\setminus\Gamma_D$ is considered to be impermeable,
i.e. the pedestrians cannot penetrate the wall $\partial\Omega\setminus\Gamma_D$.

To solve the system numerically, we use the library DUNE
and rely on a 2D Finite Element method discretization (with linear Lagrange elements)
for the space variable, with implicit time-stepping.
Note that we allow only crowds of size one, i.e. $u_1$, to exit the door.
For larger group sizes the door in impenetrable.
Such groups  really need to dissociate/degrade first and then attempt to exit.
We choose constant degradation coefficients
and take as reference values $\alpha_i=0.7$ ($i\in\{1,\dots, N\}$).
\begin{figure}[htpb]
  \centerline{ \includegraphics[width=6.5cm]{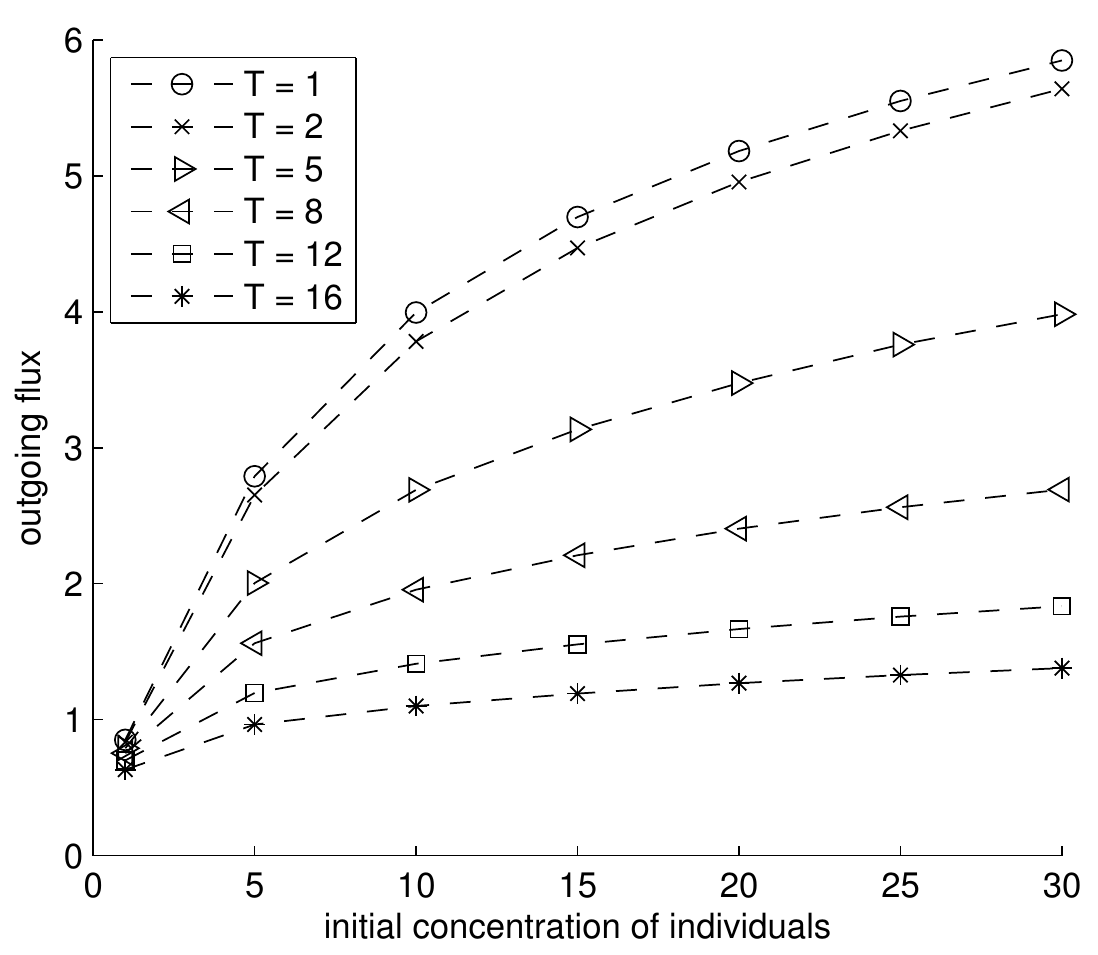}}
  \caption{Outgoing flux with respect to initial density.}
  \label{fig:thresholds}
\end{figure}

As we can see from Figure \ref{fig:thresholds}, the outgoing flux
(close to the steady state\footnote{The mass exiting the system is
evenly distribute throughout the domain $\Omega$.}) exhibits a
polynomial behavior with respect to the initial mass, where the
polynomial exponent is influenced by the choice of the threshold
$T$. It seems that the higher the threshold, the smaller is the
polynomial power.
%{\bf OLEH, is this right?? what do you mean here?} FLUX \approx M^degree{T}
This effect is rather dramatic -- it indicates that, regardless the
threshold size, behaving/moving gregariously is less efficient that
performing random walks.
\begin{figure}[htpb]
  \centerline{ \includegraphics[width=6.5cm]{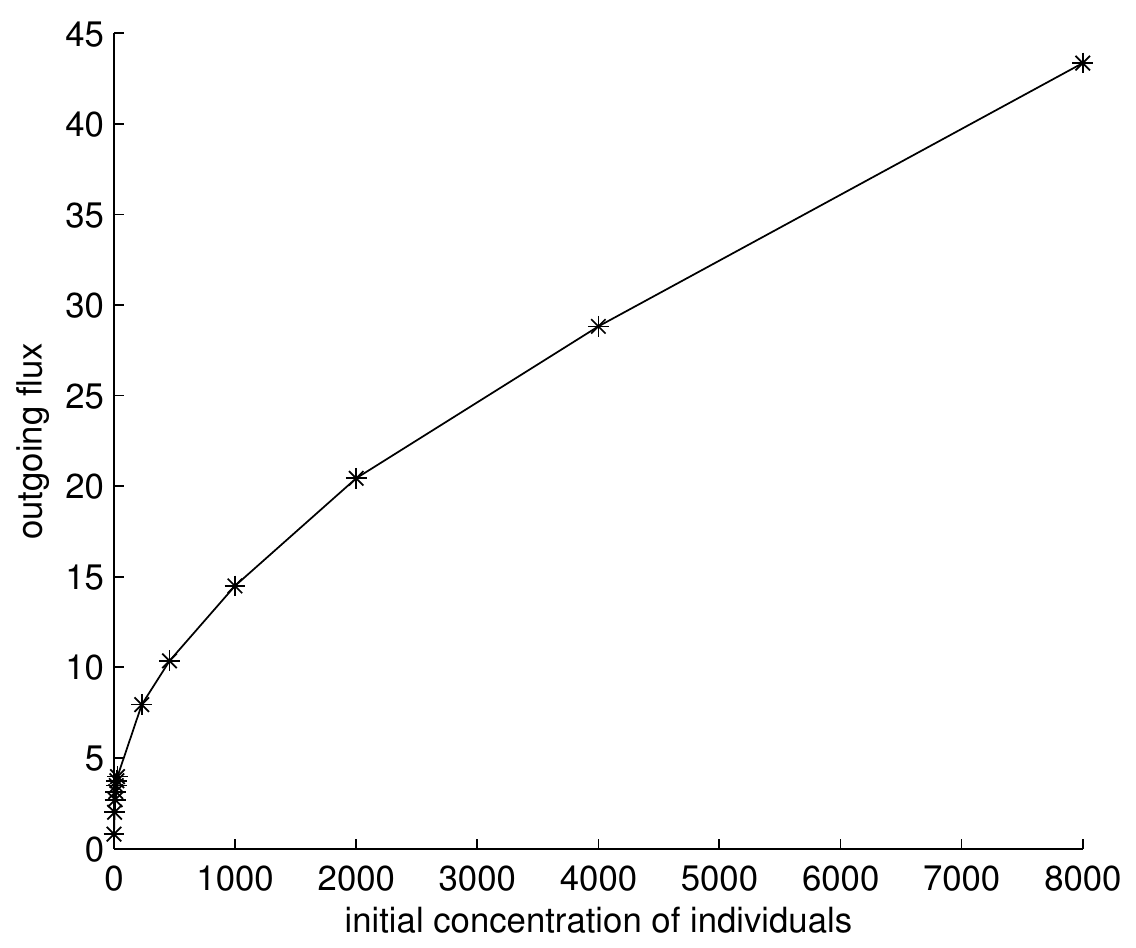}}
  \caption{Outgoing flux for $T=5$ {\em versus} large initial data $M\to \infty$.}
  \label{fig:large_mass}
\end{figure}

Figure \ref{fig:large_mass} shows that there's no apparent saturation
for the outgoing flux with respect to the mass: the growth goes on in
a polynomial fashion. The linear behavior has been obtained by setting
to zero the aggregation and degradation coefficients.

\begin{figure}[htpb]
  \centerline{ \includegraphics[width=6.5cm]{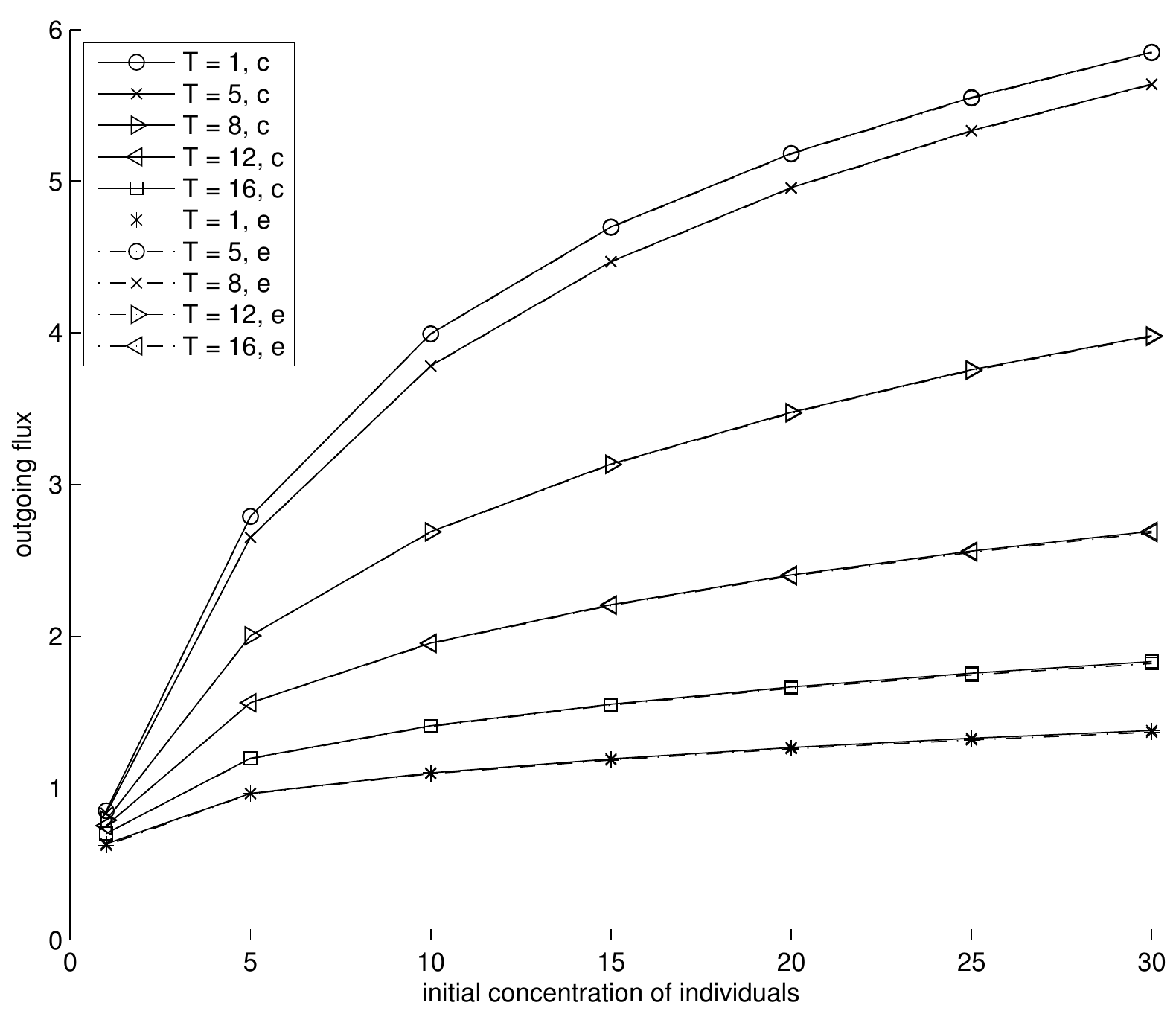}}
  \caption{Homogeneous diffusion(c) and Stokes-Einstein diffusion(e). Note that the profiles are overlapping very closely.}
  \label{fig:diffusion}
\end{figure}

In Figure \ref{fig:diffusion}, we see that the influence of variable
diffusion coefficients is marginal; since a lot of mass exchange is
happening in terms of species $u_1$, setting all the other
coefficients $d_2,\ldots,d_N$ to be lower than $d_1=1$ (i.e. bigger
groups move somewhat slower than individuals) does not affect the
output too much. Probably, the effect of diffusion could be stronger
as soon as the effective diffusion coefficients are allowed to
degenerate with locally vanishing $u_i$; this is a situation that can
be foreseen in a modified setting \cite{Varadhan}.

\begin{figure}[htpb]
  \begin{tabular}{ll}
    \includegraphics[width=180pt]{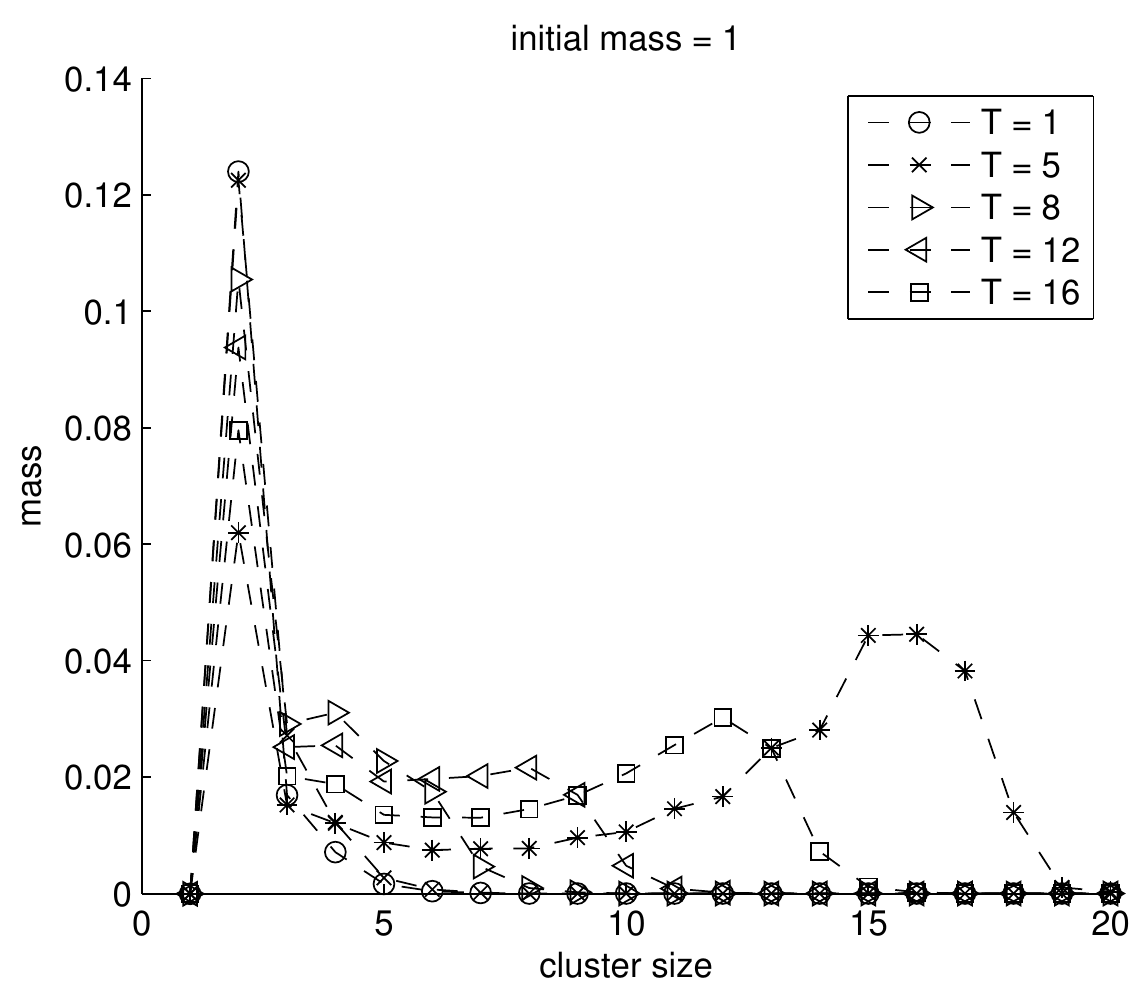}&
    \includegraphics[width=180pt]{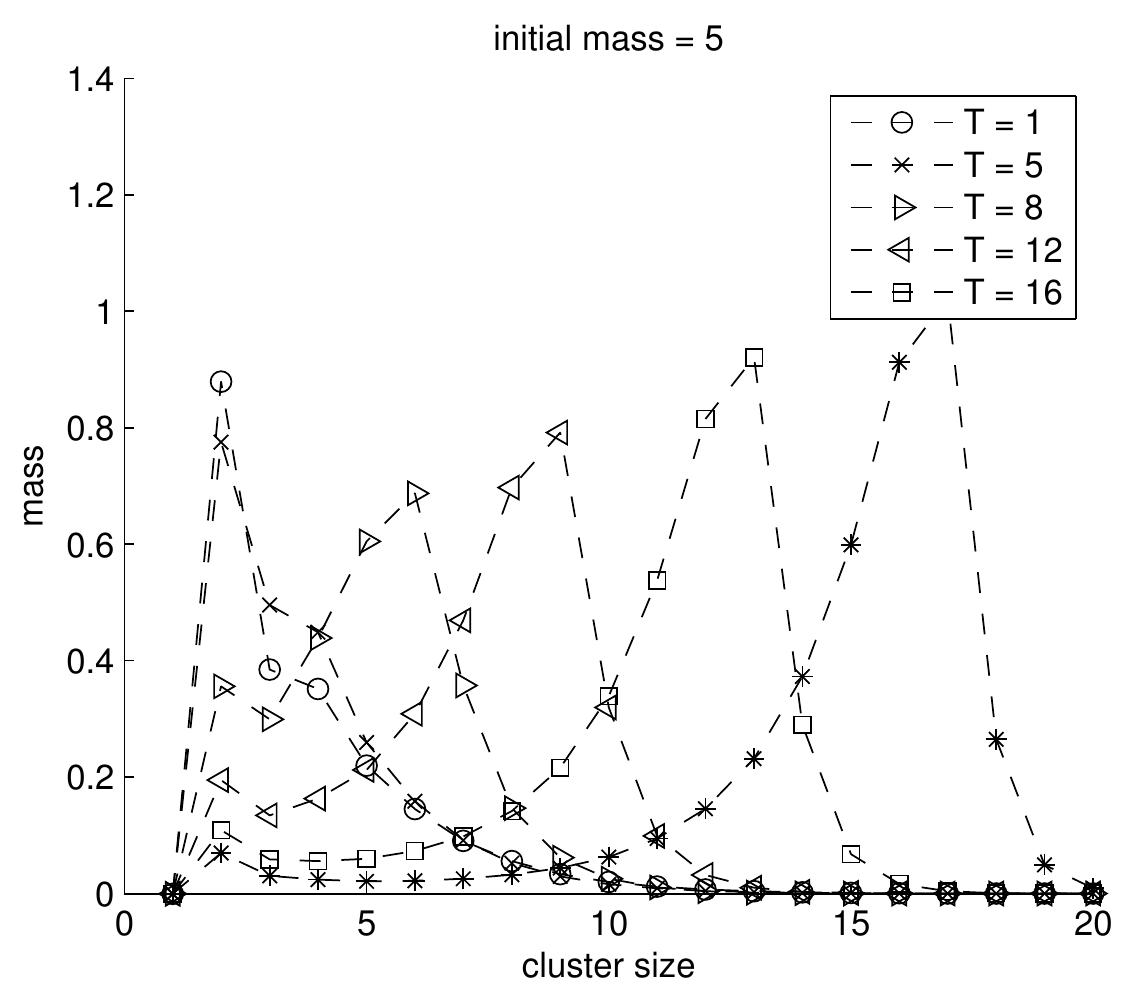}\\
    \includegraphics[width=180pt]{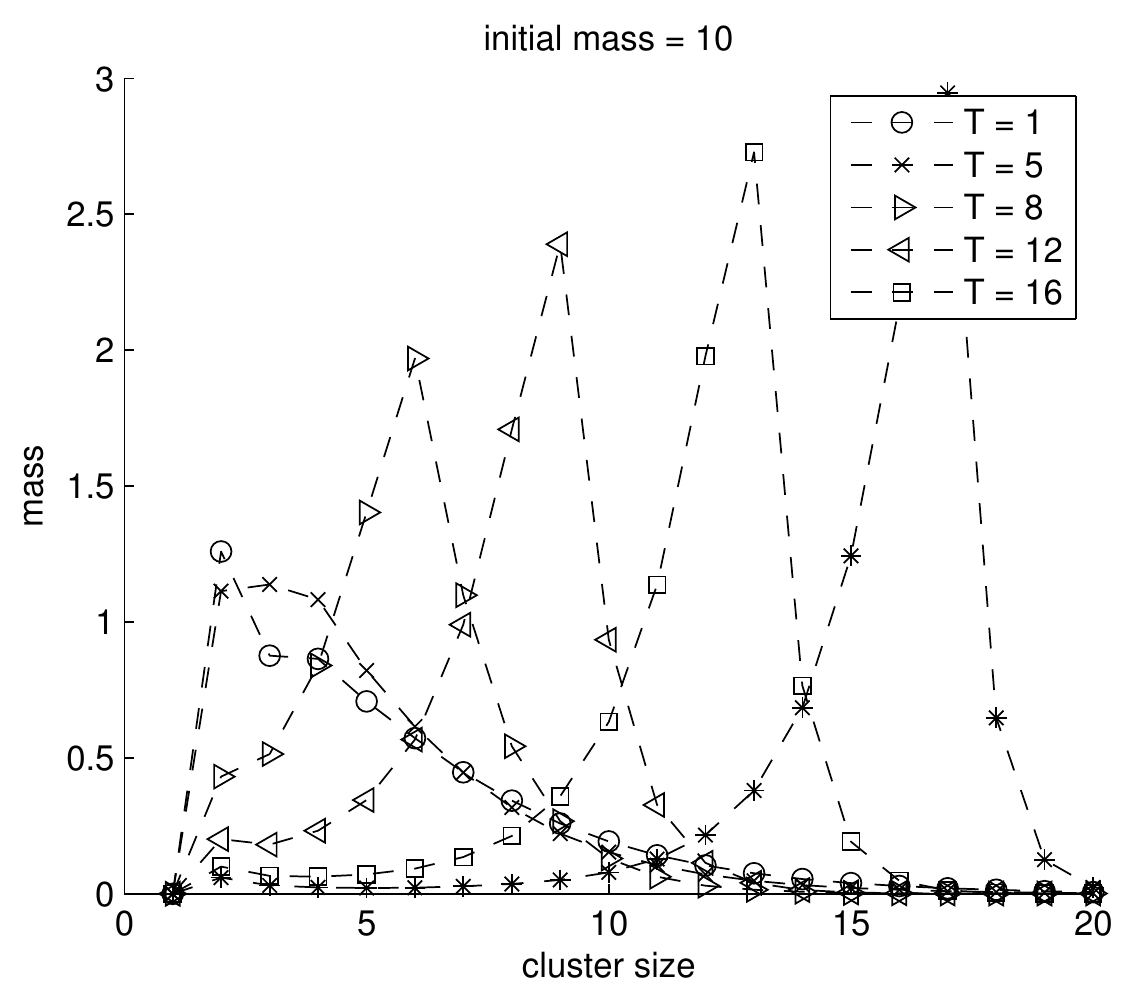}&
    \includegraphics[width=180pt]{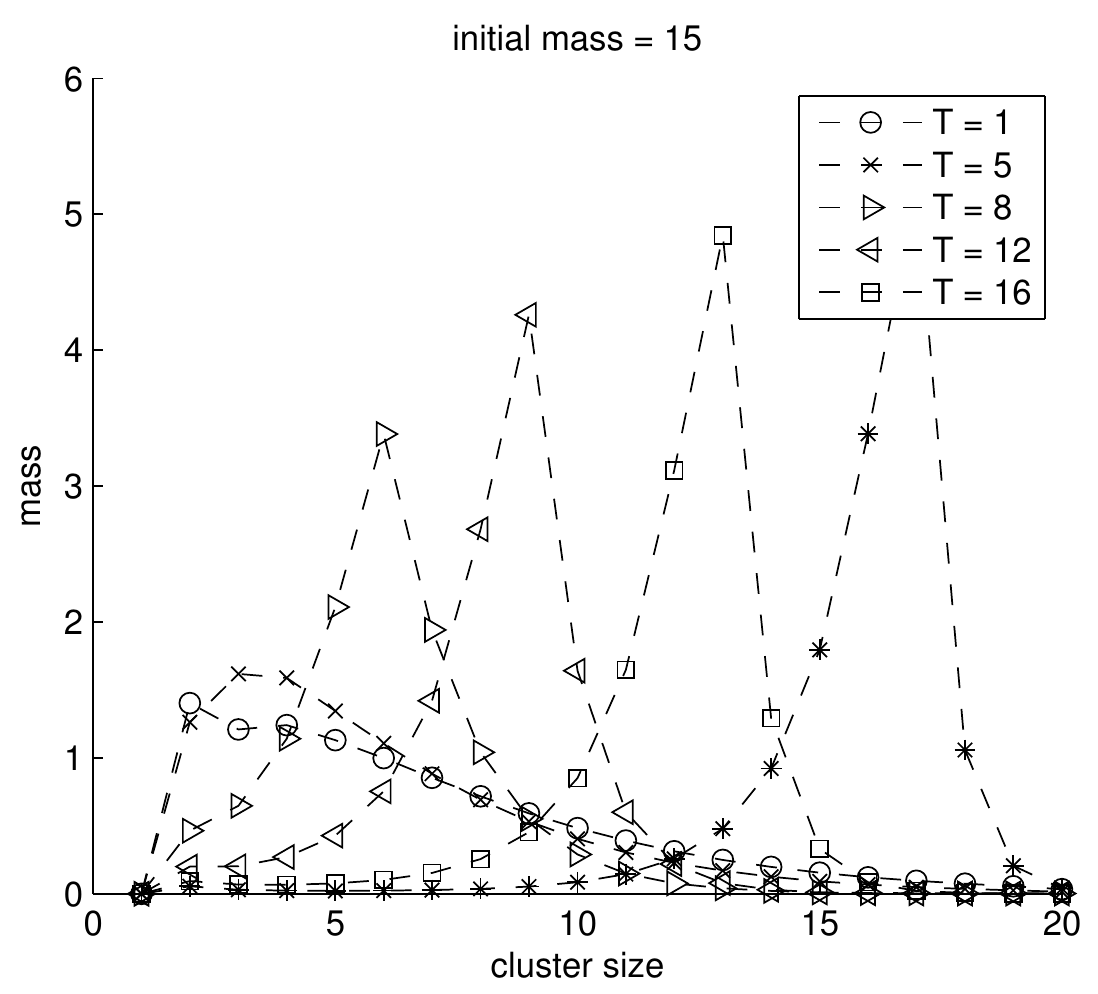}\\
    \includegraphics[width=180pt]{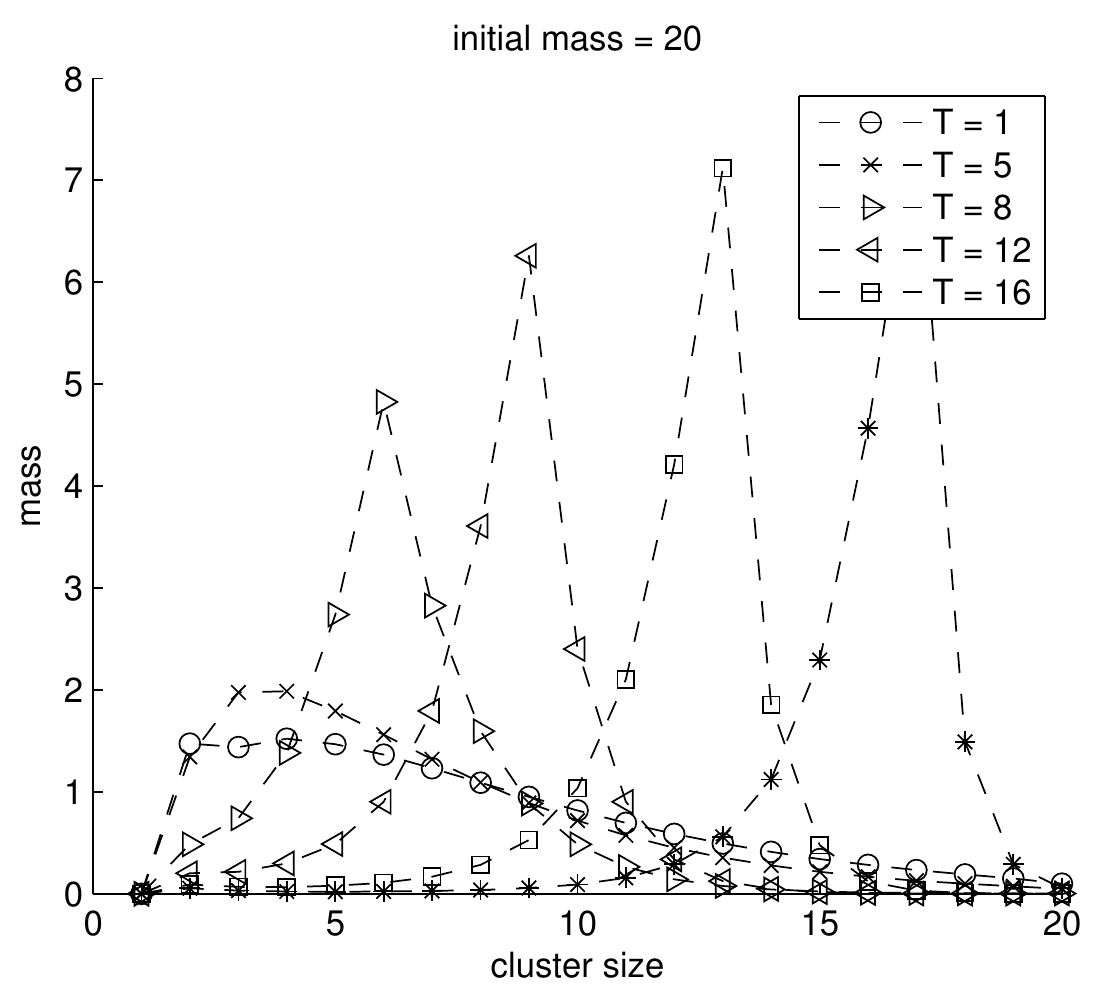}&
    \includegraphics[width=180pt]{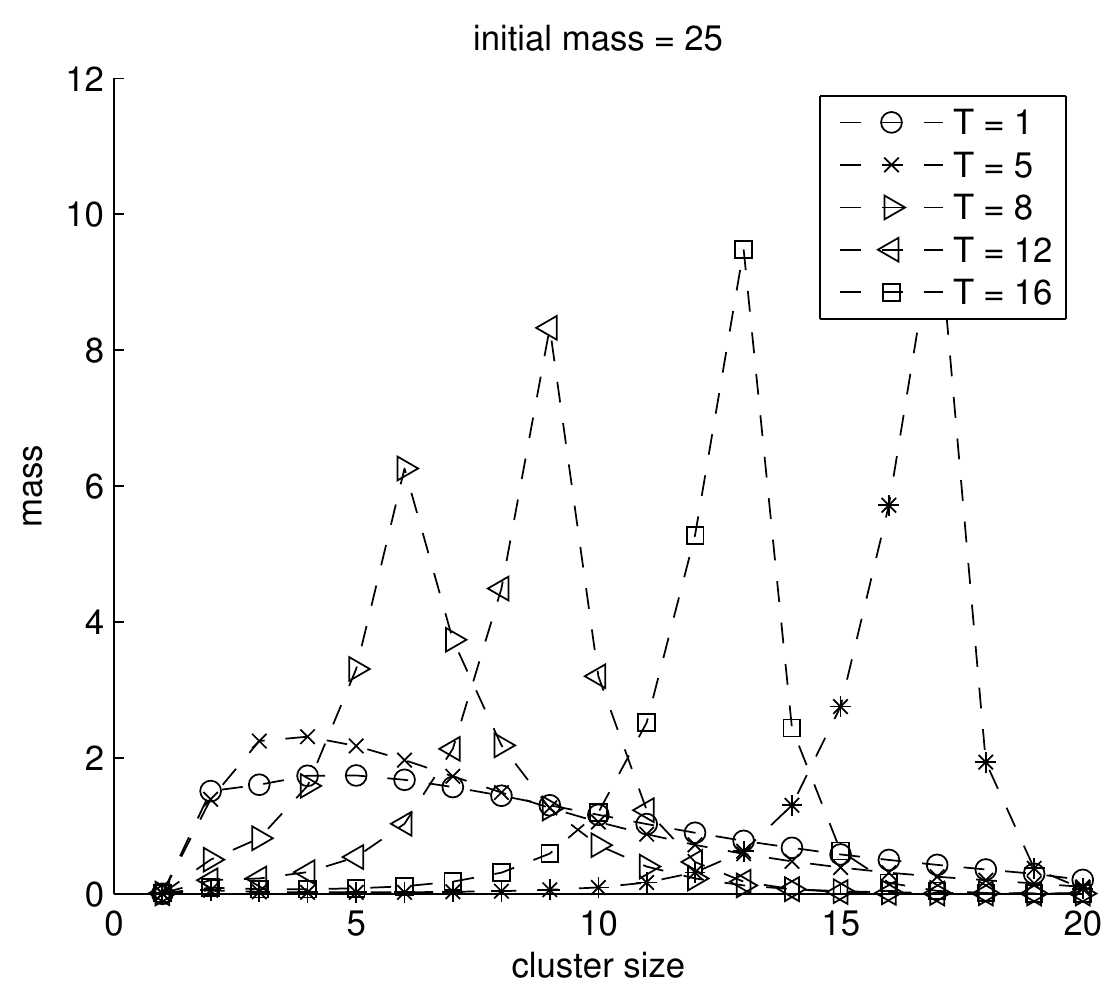}\\
    \includegraphics[width=180pt]{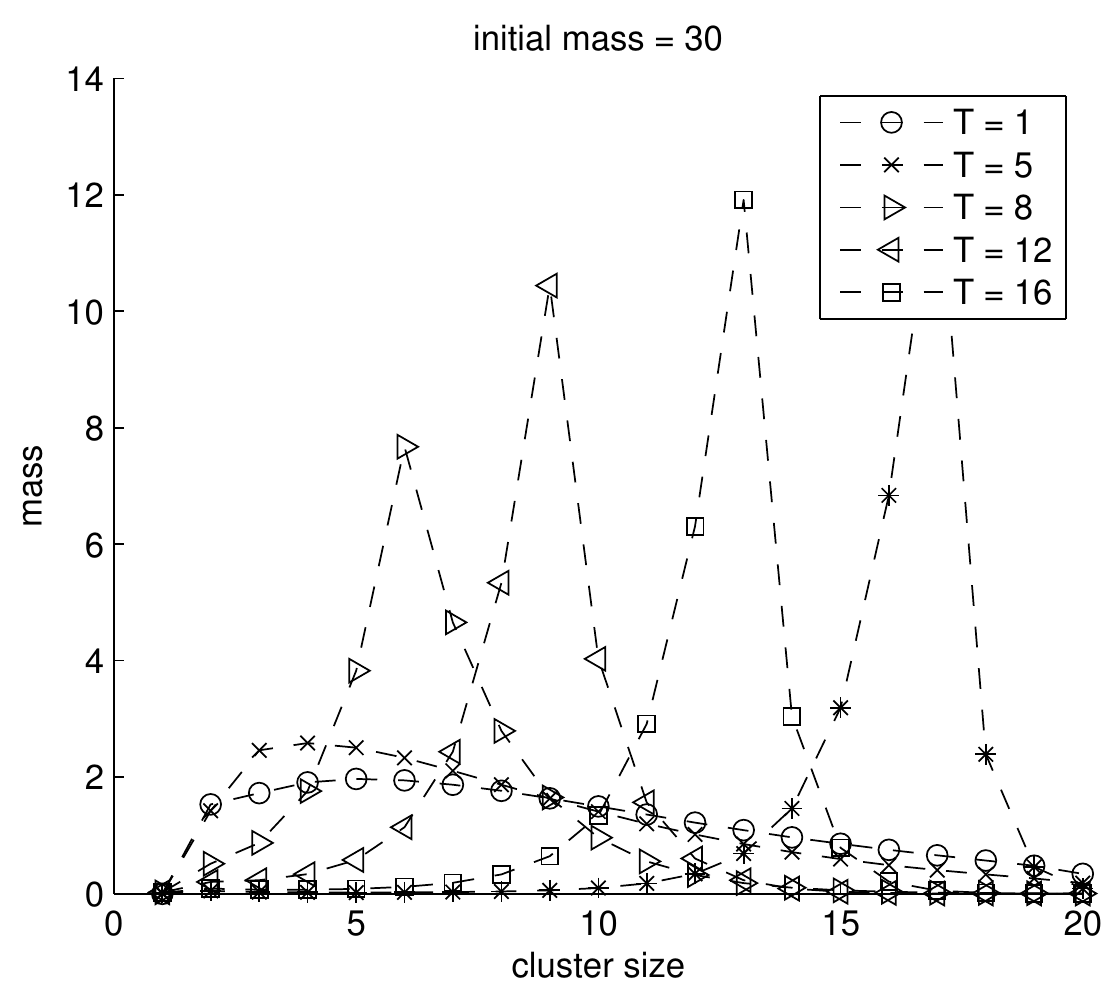}\\
  \end{tabular}
  \caption{Steady-state mass distributions. Pile-up effect around group size $T$.}\label{SSmass}
\end{figure}
\newpage
\begin{figure}[htpb]
  \begin{tabular}{ll}
    \includegraphics[width=150pt]{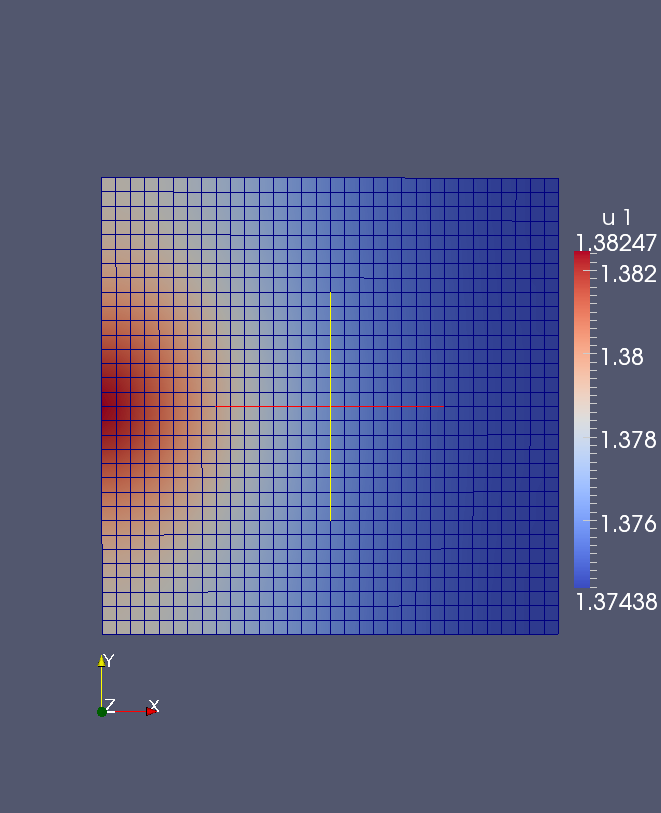}&
    \includegraphics[width=150pt]{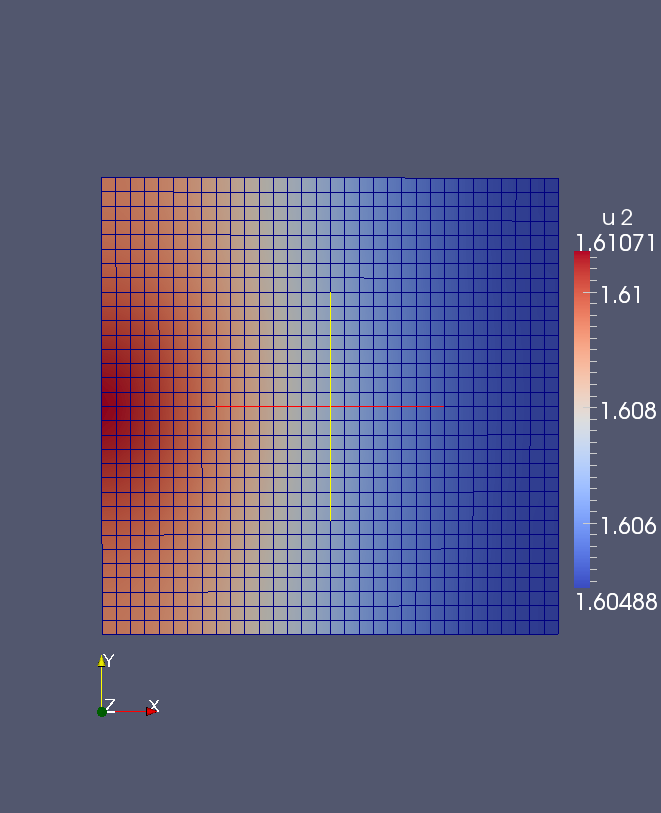}\\
    \includegraphics[width=150pt]{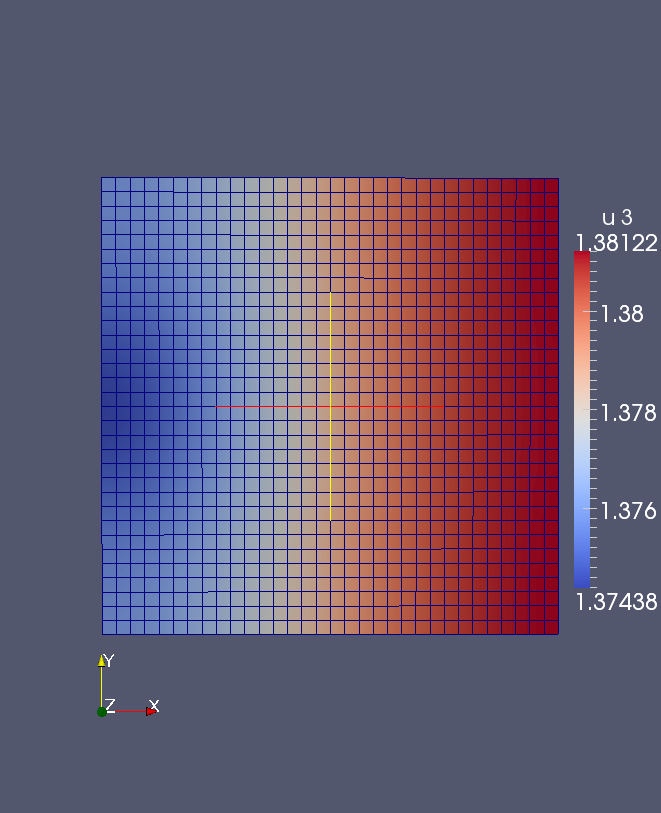}&
    \includegraphics[width=150pt]{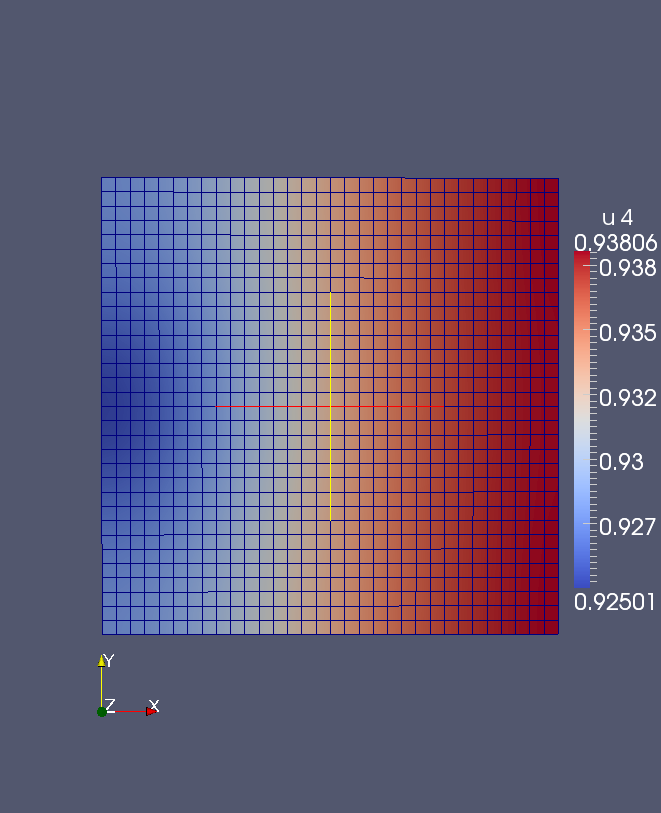}\\
  \end{tabular}
  \caption{Clusters behavior close to the exit. The case of $u_1$--$u_4$.}\label{Cluster_exit}
\end{figure}
In Figure \ref{Cluster_exit}, we see the mass escaping from the
clusters $u_1$--$u_4$ in the neighborhood of the exit. Note the
dramatic change in $u_1$ compared to what happens with the other group
sizes. It is visible that large group have to stay in the queue until
the small groups exit.

\begin{figure}[htpb]
  \centerline{ \includegraphics[width=9cm]{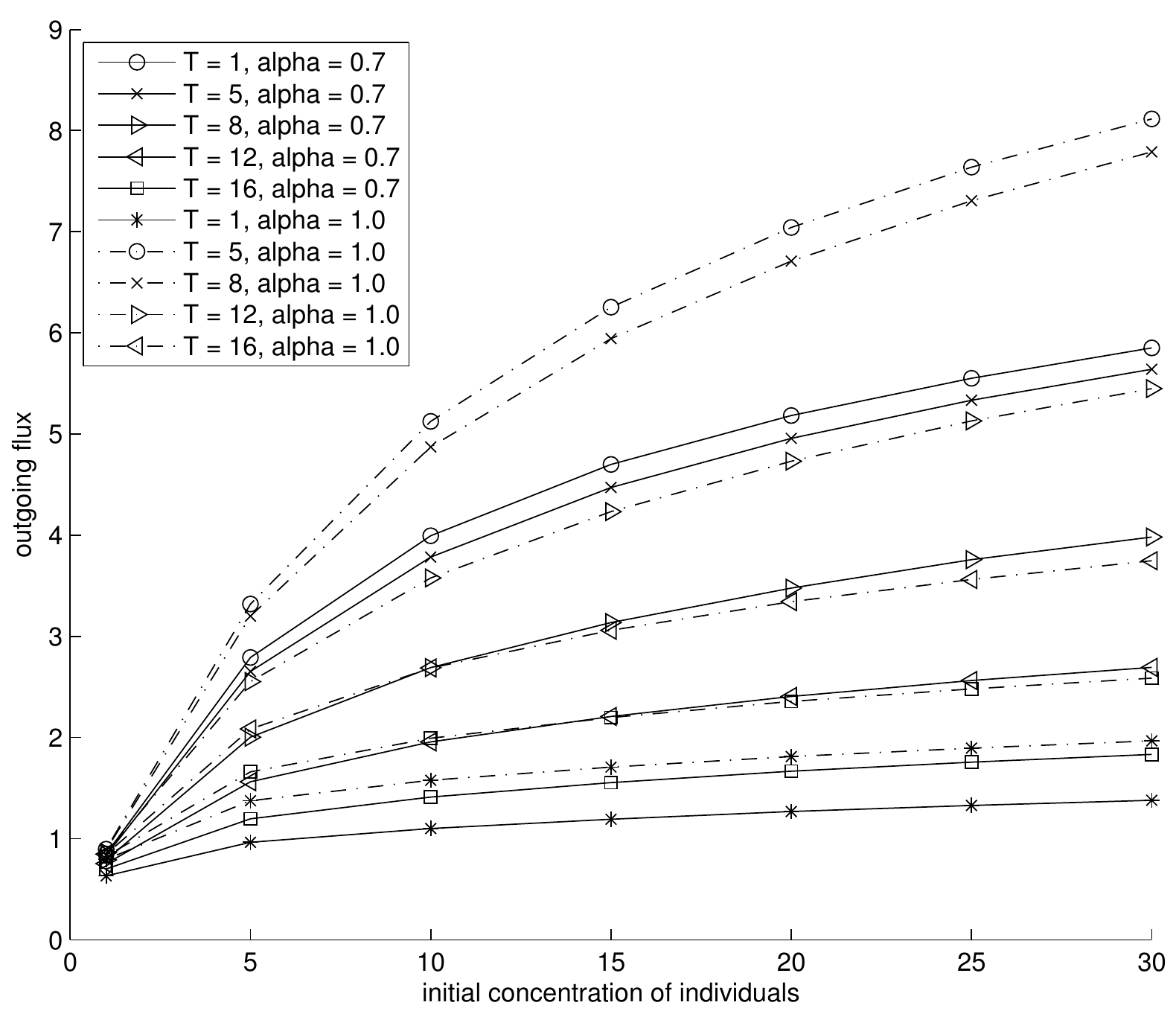}}
  \caption{Comparison of outgoing flux for different values of degradation coefficients $\alpha$.}
  \label{fig:alpha-comparison}
\end{figure}

On the other hand, we can see in Figure \ref{fig:alpha-comparison} how
the crowd breakage directly influences the outward flux. Essentially,
a faster splitting of the groups tends to increase the averaged
outgoing (evacuation) flux.  This effect is due to our choice of
boundary conditions at the exit.
We mentioned in Section \ref{BDa} that we expect that the way the
threshold $T$ intervenes in the definition of the aggregation
coefficients $\beta_i$ (compare (\ref{defT})) essentially affects the
maximum allowable group size.  We can now see that close to the steady
state situation, such situation happens. This effect is pointed out in
Figure \ref{SSmass}; the picture suggests that the mass of pedestrians
piles-up in structures whose maximum lie around $T$.
% \newpage

\CCLsection{A lattice model for the reverse {\em mosca cieca} game}
\label{s:random}
\CCLsubsection{Microscopic dynamics}\label{mosca}
Using the lattice model presented in this section, we explore the
effects of the microscopic non--exclusion on the overall exit flux
(evacuation rate). More precisely, we look again at social thresholds
and study this time the effect of the {\em buddying threshold} (of
no--exclusion per site) on the dynamics of the crowd and investigate
to which extent such approach confirms the following pattern revealed
by investigations on real emergencies and also emphasized in Section
\ref{BDa}:
{\em
  If the evacuees tend to cooperate and act altruistically, then their
  collective action tends to favor the occurrence of
  disasters}\footnote{Note that,due to the lack of visibility,
    anticipation effects (see \cite{Suma}) and drifts (see \cite{Guo}) are
    expected to play no role in evacuation.}.

% We ask ourselves to which extent our model confirms the following
% pattern revealed by investigations on real emergencies: If the
% evacuees tend to cooperate and act altruistically, then their
% collective action tends to favor the occurrence of
% disasters.\footnote{Note that due to the lack of visibility
% anticipation effects (see \cite{Suma}) and drifts (see \cite{Guo}) are
% expected to play no role in evacuation.}

Question (Q1) in Paragraph~\ref{introduction} drives any possible attempt
of modeling pedestrians motion.
In this section we show how an answer to this question can be setup
by using a stochastic point of view.

\begin{center}
  \begin{figure}[h]
    \centering
    % \begin{picture}(500,180)(-120,0)
    \includegraphics[width=9cm]{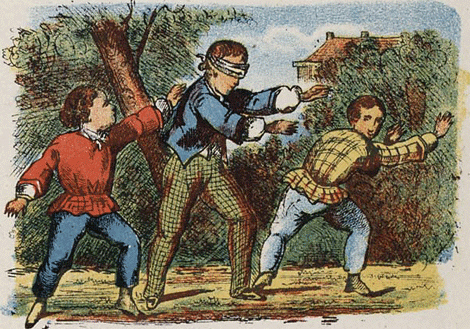}
    % \end{picture}
    \caption{The blind man's buff game  (the {\em mosca cieca} ({\em ital.}) game).}
    \label{mosca_game}
  \end{figure}
\end{center}

Our reference scenario is here a microscopic one:
Imagine to be one of the individuals in a dark (possibly
crowded) corridor trying to save your life by quickly
reaching one of the exits. You cannot see anything and, maybe,
you do not have any {\em  a priori} knowledge of the geometry of the
corridor  you have to exit from.
It is not difficult to imagine that you will not be able to
keep a constant direction of motion and that, in any case,
it will be not chosen via some neat reasoning, but you will essentially
chose it at random on the basis of what other people shout and scream.
In some sense your motion will closely resemble that of the
blinded kid playing
{\em mosca cieca}\footnote{{\em Mosca cieca}
  means in Italian {\em blind fly}. It is the Italian name of
  a traditional children's game also known as
  {\em blind man's buff} or {\em blind man's bluff}.
  The game is played in a spacious free of dangers area
  in which one player, the ``mosca", is blindfolded and moves around
  attempting to catch the other players without being able to see them.
  Other players try to avoid him; they
  make fun of the ``mosca" inducing him to change direction.
  When one of the player is finally caught,  the ``mosca" has to identify
  him by touching is face
  and if the person is correctly identified he becomes the
  ``mosca". Interestingly, the game has inspired significantly satiric
  literature \citep{Manzoni,Musatescu,bogdanov2001game}.
  Our model tackles  a reverse mosca cieca game -- all the
  players (pedestrians) cluster around, as if they were
  blindfolded, trying to catch the (invisible) exit.
  Note that the game is actually international
  {\em  \begin{otherlanguage}{russian}
      жмурки
    \end{otherlanguage}}
  (Russian),
  % ??????
  {\em baba-oarba} (Romanian), {\em Blindekuh} (German) ...
}\footnote{The picture in Figure \ref{mosca_game} is taken from \\ http://commons.wikimedia.org/wiki/File:Jongensspelen$\_$14.jpg.}
with his friends.

This simple remark triggered us to propose a stochastic model
for the pedestrian motion in no--visibility areas based
on a random walk scheme \cite{CirilloMuntean2012,CirilloMuntean2012crm}.
The random walk rule has been introduced by taking into account
a possible interaction between the individuals, see the
question (Q2) in Section~\ref{introduction}.

Pedestrians move freely inside the corridor and  like to buddy with people
they accidentally meet at a certain point (site). The more people are
localized at a certain site, the stronger the preference to attach to it.
However if the number of people at a site reaches a threshold, then such
site becomes not attracting
for eventually new incomers.

Our lattice model provides a not so nice answer: In many situations,
it seems much better not to cooperate\footnote{"Cooperation" means in
this setting "buddying" - the basic gregarious tendency. Our current
modeling approach does not yet allow the particles to influence each
other. We refer the reader to \cite{Eggels} for a setting where
particles do exchange mass (as a measure of "confidence") not only
momentum.}. More precisely, in Section \ref{play}, we will see that
simulations indicate to
\begin{itemize}
\item[--]
  cooperate with one person at time;
\item[--]
  cooperate with more than one person only if the number of evacuees in
  the corridor is not too large.
\end{itemize}

Based on this idea we have announced in \cite{CirilloMuntean2012crm}
and then presented in details in \cite{CirilloMuntean2012} a
model\footnote{The model proposed in the paper is slightly more
complicated, for instance there it is taken into account the
possibility to tune the interaction between the pedestrians and the
wall of the corridor} for the motion of pedestrians governed by the
following four mechanisms:
\begin{itemize}
\item[(A1)]
  in the core of the corridor,
  people move freely without constraints;
\item[(A2)]
  the boundary is reflecting;
\item[(A3)]
  people are attracted by bunches of other people up
  to a threshold
  ({\em buddying mechanism});
\item[(A4)]
  people are blind in the sense that there is no drift
  (desired velocity) leading them towards the exit.
\end{itemize}

Let $\Lambda\subset\mathbb{Z}^2$ be a finite square with
odd side length $L$. We refer to this as  the {\em corridor}.
Each element $x$ of $\Lambda$ will be called a {\em cell} or {\em site}.
The external boundary of the corridor is made of four segments
made of $L$ cells each;
the point at the center of one of these four sides is called {\em exit}.
Let $N$ be  positive integer denoting the (total)
{\em number of individuals} inside the corridor $\Lambda$.
We consider the state space $X:=\{0,\dots,N\}^\Lambda$.
For any state $n\in X$, we let
$n(x)$ be the {\em number of individuals} at cell $x$.

We define a Markov chain $n_t$ on the finite state space
$X$ with discrete time $t=0,1,\dots$.
The parameter of the process is the integer (possibly equal
to zero) $T\ge0$ called
{\em threshold}.
We finally define the function $S:\mathbb{N}\to\mathbb{N}$
such that
\begin{displaymath}
  S(k):=
  \left\{
    \begin{array}{ll}
      1 & \textrm{ if } k>T\\
      k+1 & \textrm{ if } k\le T\\
    \end{array}
  \right.
\end{displaymath}
for any $k\in\mathbb{N}$. Note that for $k=0$ we have $S(0)=1$.

\begin{figure}[htp]
  \begin{center}
    \begin{picture}(500,90)(-15,0)
      \thicklines
      \put(0,0){\line(1,0){20}}
      \put(0,0){\line(0,1){20}}
      \put(0,20){\line(1,0){20}}
      \put(20,0){\line(0,1){8}}
      \put(20,20){\line(0,-1){8}}
      \put(8,9){\circle*{3}}
      \put(27,7){in the bulk}

      \put(200,0){\line(1,0){20}}
      \put(200,0){\line(0,1){20}}
      \put(200,20){\line(1,0){20}}
      \put(220,0){\line(0,1){8}}
      \put(220,20){\line(0,-1){8}}
      \put(203,17){\circle*{3}}
      \put(227,7){in a corner}

      \put(0,50){\line(1,0){20}}
      \put(0,50){\line(0,1){20}}
      \put(0,70){\line(1,0){20}}
      \put(20,50){\line(0,1){8}}
      \put(20,70){\line(0,-1){8}}
      \put(8,53){\circle*{3}}
      \put(27,57){close to the boundary}

      \put(200,50){\line(1,0){20}}
      \put(200,50){\line(0,1){20}}
      \put(200,70){\line(1,0){20}}
      \put(220,50){\line(0,1){8}}
      \put(220,70){\line(0,-1){8}}
      \put(217,60){\circle*{3}}
      \put(227,57){in front of the exit}
    \end{picture}
    \caption{Schematic description of the different situation
      considered in the definition of the transition matrix.}
    \label{f:salti}
  \end{center}
\end{figure}
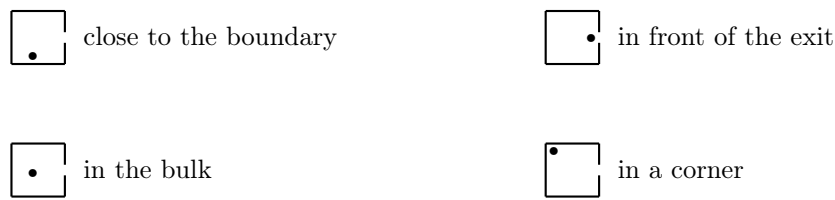

The transition matrix of the Markov chain is specified by
assigning the stochastic
rule according to which the individuals move on the lattice.
At each time $t$, the $N$ individuals move simultaneously within the corridor
according to the rules that will be specified in the following. These
rules depend on the location of the pedestrian, we have to distinguish
among four cases: bulk, corner, neighboring the wall, and
neighboring the exit (see Figure~\ref{f:salti}.
In the bulk: the probability for a pedestrian at the site $x$ to
jump to one of the four neighboring sites $y_1,\dots,y_4$ is
\begin{displaymath}
  \frac{S(n(y))}{S(n(x))+S(n(y_1))+\cdots+S(n(y_4))}.
\end{displaymath}
In a corner: the probability for a pedestrian at the site $x$ to
jump to one of the two neighboring sites $y_1$ and $y_2$ is
\begin{displaymath}
  \frac{S(n(y))}{S(n(x))+S(n(y_1))+S(n(y_2))}.
\end{displaymath}
In a site close to the boundary: the probability for a pedestrian at the
site $x$ to jump to one of the three neighboring sites $y_1$, $y_2$,
and $y_3$ is
\begin{displaymath}
  \frac{S(n(y))}{S(n(x))+S(n(y_1))+S(n(y_2))+S(n(y_3))}.
\end{displaymath}
In front of the exit: the probability for a pedestrian at the
site $x$ to jump to one of the three neighboring sites $y_1$, $y_2$, and
$y_3$ in the bulk is
\begin{displaymath}
  \frac{S(n(y))}{S(n(x))+S(n(y_1))+S(n(y_2))+S(n(y_3))+(T+1)},
\end{displaymath}
whereas the probability to exit is
\begin{displaymath}
  \frac{T+1}{S(n(x))+S(n(y_1))+S(n(y_2))+S(n(y_3))+(T+1)}.
\end{displaymath}
In all the cases described above, the probability for the
individual to stay at the same site $x$ (not to move)
is $S(n(x))$ divided by the corresponding normalization
denominator.

%% Fig. f:salti

The dynamics is then defined as follows:
at each time $t$,  the position of all the individuals
on each cell is updated according to the probabilities defined
above. If one of the individuals jumps on the exit
cell a new individual is put  on a cell of $\Lambda$
chosen randomly with the uniform probability $1/L^2$.

\CCLsubsection{Playing games on lattices}\label{play}

The possible choices for the parameter $T$ correspond to
two different physical situations.
For $T=0$ the function $S(k)$ is equal to one
whatever the occupation numbers. This means that
each individual has the same probability to jump to one of
its nearest neighbors or to stay on his site.
This is the independent symmetric random walk case with not zero resting
probability.
The second physical case is $T>0$. For instance,
$T=1$ means mild buddying, while
$T=100$ would express an extreme buddying.
No simple exclusion is included in this model:
on each site one can cluster as many particles (pedestrians) as one wants.
The basic role of the threshold is the following:
The weight associated to the jump towards the site $x$
increases from $1$ to $1+T$ proportionally to the occupation
number $n(x)$ until $n(x)=T$, after that level it drops back to $1$.
Note that this rule is given on weights and not to probabilities. Therefore,
if one has $T$ particles at $y$ and $T$ at each of its nearest
neighbors, then  at the very end one will have that the probability to stay
or to jump to any of the nearest neighbors is the same. Differences
in probability are seen only if one of the five (sitting in the core) sites
involved in the jump (or some of them) has an occupation number
large (but smaller than the threshold).

In \cite{CirilloMuntean2012},
we have studied numerically this model
for $T=1,2,5,30$, and $100$.
The Monte Carlo simulations have been all performed for $L=101$.
For each value of the
threshold we have studied the cases
$N=100,600,1000,6000,10000$.
For the choices $T=30$ and $T=100$ we have also analyzed the
cases
$N=2000,2200,2400,2600,2800,3000,3300$
and
$N=1300,1600,2000,3000$,
respectively.

The main quantity of interest that one has to compute
is the {\em average outgoing flux}
that is to say the ratio between the number of
individuals which exited the corridor in the time interval $[0, t_\rr{f}]$
and $t_\rr{f}$.
This quantity fluctuates in time,
but for times large enough it approaches a
constant value. In order to observe relative fluctuations smaller than
$10^{-2}$ we had to use $t_\rr{f}=5\times10^{6}$. To capture  the extreme
buddying case $T=100$, we used $t_\rr{f}=1.5\times10^{7}$.

\begin{figure}[t]
  \begin{center}
    \includegraphics[angle=-90,width=0.7\textwidth]{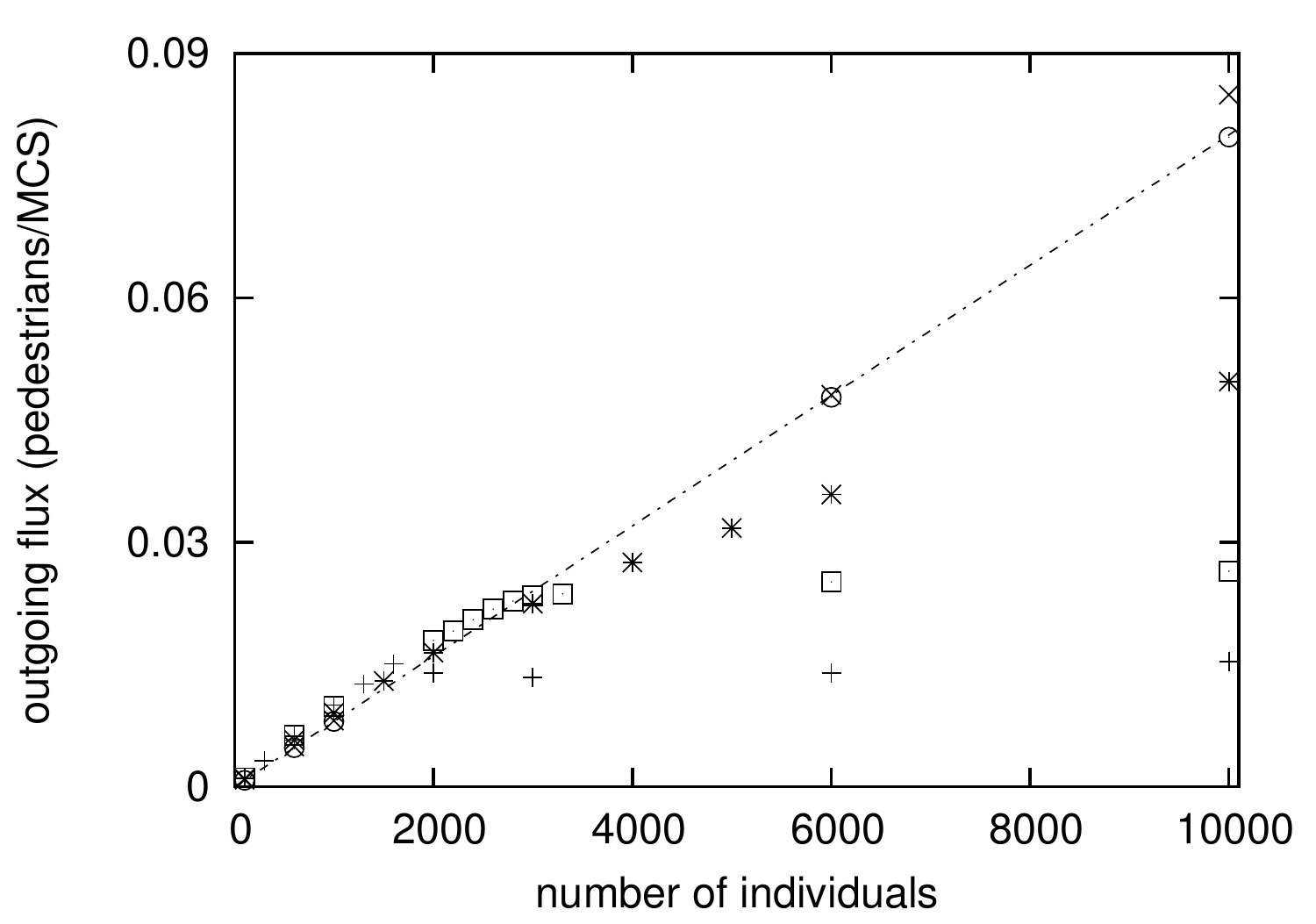}\\
    \caption{Averaged outgoing flux vs. number of pedestrians.
      The symbols $\circ$, $\times$, $*$, $\square$, and $+$
      refer respectively to the cases $T=0,1,5,30,100$.
      The straight line has slope $8\times10^{-6}$ and has
      been obtained by fitting the Monte Carlo
      data corresponding to the case $T=0$.}
    \label{f:due}
  \end{center}
\end{figure}

Figure~\ref{f:due} depicts our results, where the averaged outgoing flux
is given as a function of the
number of individuals.
At $T=0$, that is when no buddying between the individuals is
put into the model, the outgoing flux results proportional to
the number of pedestrians in the corridor;
indeed the data represented by the symbol $\circ$ in Figure~\ref{f:due}
have been perfectly fitted by a straight line.

The appearance of the straight line was expected in the case $T=0$ since
in this case the dynamics reduces to that of a simple symmetric
random walk with reflecting boundary conditions; see also the straight
line in Figure \ref{fig:thresholds} (where we suspect that,
microscopically, something very similar microscopically happens).
This effect was studied rigorously in the one--dimensional case and via Monte Carlo
simulations in dimension two in \cite{ABCM2011}.
The order of magnitude of the
slope can be guessed with a simple argument \cite{ABCM2012}:
the typical time needed by the walker, started
at random in the lattice, to reach the site facing the
exit is of order of
\begin{displaymath}
  \Big(\frac{1}{6}L\Big)^2
  \times 4L
  =
  \frac{1}{9}L^3.
\end{displaymath}
The first term is the square of the
average distance of a point
inside a square of side length $L$
from the boundary of the square itself
and the second one is
the number of times the walker has to visit
the internal boundary before facing the exit.
Hence
\begin{displaymath}
  \textrm{outgoing flux}
  =
  \frac{1}{t_{\textrm{f}}}N
  \frac{t_{\textrm{f}}}{L^3/9}
  =
  \frac{9}{L^3}\,N
  = 8.73\times10^{-6}\,N.
\end{displaymath}

When a weak buddying effect is introduced in the model, that is in the
case $T=1$, we find that if the number of individuals is small enough,
say
$N\le6000$, the behavior is similar to the one measured in the absence of
buddying ($T=0$). At $N=10000$, on the other hand,
we measure a larger flux; meaning that in the {\em crowded} regime
small buddying favors the evacuation of the corridor
[i.e. it favors the finding of the door].

The picture changes completely when buddying is increased. To this end,
see the cases $T=5,30,100$.
The outgoing flux is slightly favored when the number of individuals
is low and strongly depressed when it this becomes high.
The value of $N$ at which this behavior changes strongly depends
on the threshold parameter $T$.

\begin{figure}[t]
  \begin{center}
    \includegraphics[angle=-90,width=0.7\textwidth]{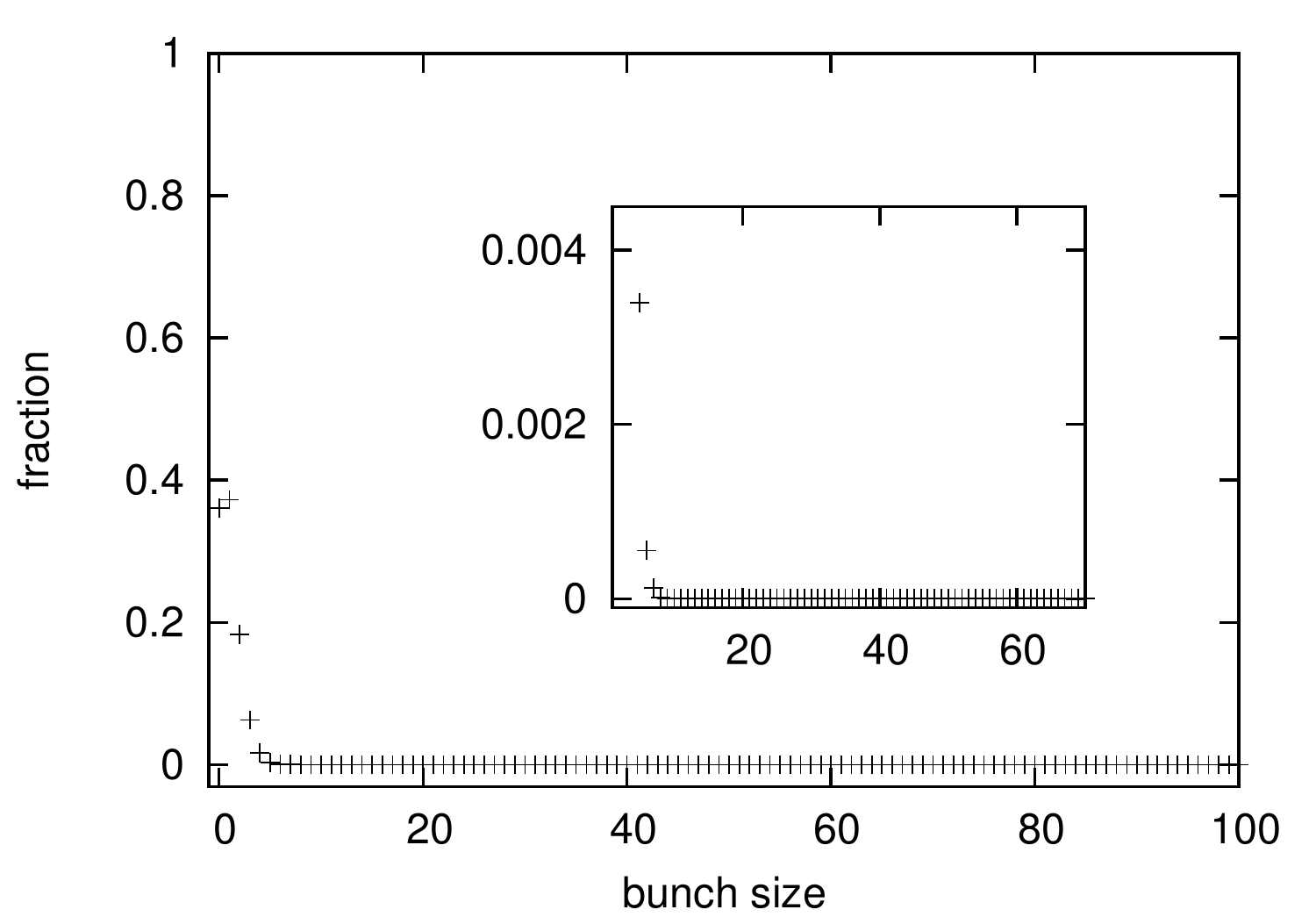}\hfill
    \includegraphics[angle=-90,width=0.7\textwidth]{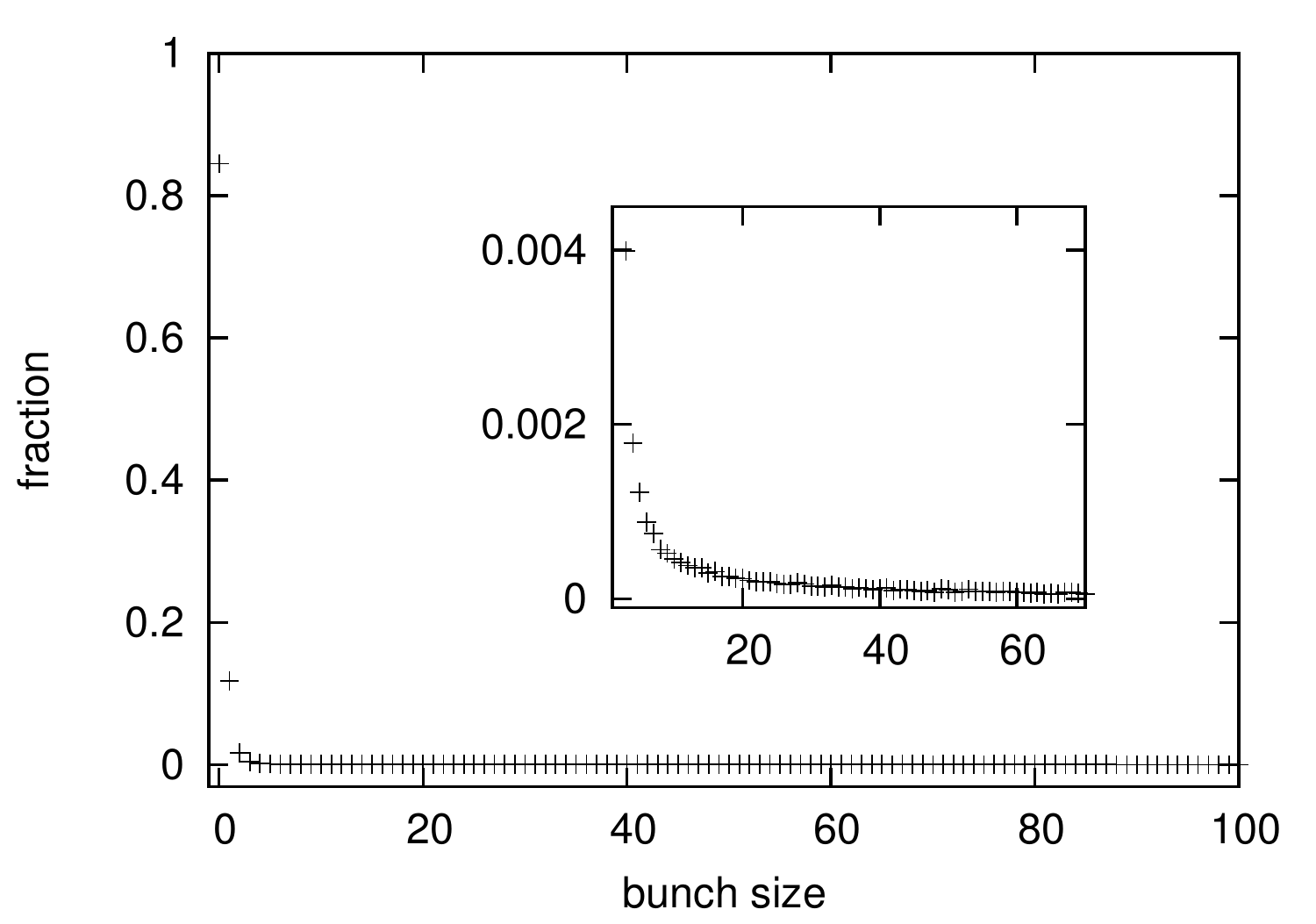}
    \caption{Histogram of the size of the bunch of people occupying
      the center of the lattice for $N=10000$, $T=0$ (left),
      and $T=100$ (right).}
  \end{center}
  \label{f:bunch}
\end{figure}

The question remains:\begin{center}
  \fbox{Why does the disaster occur at large threshold and large density?}
\end{center}
It is not straightforward to understand how the model behaves in this regime.
Inspired by theory behind particles percolation in porous media, one
possible natural explanation would be that individuals cluster in
bunches and that the resulting dynamics is characterized by the motion
of these huge groups.
At the moment we do not know if this explanation is the right one. In
order to support it at least partially,  we have computed the histogram of the size of the
bunch at the center of the corridor; see Figure~\ref{f:bunch}.
Here we compare the cases $T=0$ and $T=100$ for
$N=10000$ individuals.
The histogram has been constructed by running a $10^6$ long
simulation.
The picture does suggest that
the bunch formation is negligible in the former case while in the latter
it is a possible mechanism.

% \begin{figure}[t]
%   \begin{center}
%     \begin{picture}(500,160)
%       \put(-15,160)
%       {
%       \resizebox{8cm}{!}{\rotatebox{-90}{\includegraphics{fig-bunch0.eps}}}
%     }
%       \put(225,160)
%       {
%       \resizebox{8cm}{!}{\rotatebox{-90}{\includegraphics{fig-bunch100.eps}}}
%     }
%     \end{picture}
%     \caption{Histogram of the size of the bunch of people occupying
%     the center of the lattice for $N=10000$, $T=0$ (left),
%     and $T=100$ (right).}
%     \label{f:bunch}
%   \end{center}
% \end{figure}

Now, we can summarize our conclusions based on this microscopic model.
Through a novel lattice model we have examined the effect of buddying mechanisms
on the efficiency of evacuation in a smoky corridor (no--visibility
area).
With respect to the outgoing flux measured in absence
of group formation, our model predicts that
\begin{itemize}
\item[--]
  the existence of many small groups (threshold $T$ equal to one)
  favors the exit efficiency
  (compare points and straight line
  in figure~\ref{f:due}: straight line is essentially the not--buddying case);
\item[--]
  strong gregariousness  favors the exit efficiency only if the
  number of evacuees is small enough;
\item[--]
  the larger the threshold, the more dramatic is this effect.
\end{itemize}

In \cite{Simo}, the authors present an experiment whose purpose was to
study evacuees exit selection under different behavioral objectives.
The evacuation
(egress) time of the whole crowd turned out to be shorter when the evacuees
behave egoistically instead of behaving cooperatively. This is rather
intriguing and counter intuitive fact, and it is very much in the spirit of the
effect of the threshold $T$ we observed above.

Note that for low densities the buddying mechanism increases
the outgoing flux,
whereas at large densities the scenario is dramatic:
isolated individuals may turn
to have a bigger escape chance than a large group around a leader [behavior
recommended by standard manuals on evacuation strategies, see e.g.
\cite{NIBHV}, p. 122.]. This suggests that evacuation strategies should not
rely too much only  on the presence of a leader; see \cite{Kats} for a related scenario.

\CCLsection{Open issues}
\label{s:open}
This research opens a series of fundamental questions. Some of them
connect to the psychology of pedestrian groups that are essentially
driven by features, behaviors, and not necessarily by desired
velocities encoding the information on the location and accessibility
of the exits.  Some other questions are more general and refer to
effect of the threshold on the general behavior of solutions to both
cellular-like automata (lattice systems) as well as on
Becker-D\"oring-like systems of differential equations (continuum
systems).

We conclude the paper by enumerating a few detailed questions as well
as less crystalized but promising links to other fields of science:
\begin{enumerate}
\item[(i)] Is there a direct link between the models (or variants on
  the same theme) presented in Section \ref{BDa} and in Section
  \ref{s:random}? Can one derive in the many-particle limit (i.e. $N\to
  \infty$)) Becker-D\"oring-like equations having as departure point a
  particle system with threshold dynamics governing the interactions? We
  expect that a few hints can be taken over from \cite{Gross} at least
  in what the moderately stochastically interacting particle limit case
  is concerned. Note that some ideas on how one could possibly treat
  simple interacting-particle systems with threshold are also
  anticipated in \cite{Bodineau}, e.g., in the context of modeling
  batteries.  For the passage from the Becker-D\"oring-like system to
  the corresponding continuity equation, ideas from \cite{Barbara} may
  turn to be useful.

\item[(ii)] We do not know yet how pedestrians should behave if they
  don't posses any information on the location of the exit. Difficult
  questions are: What is the right type of behavior in the dark? or How
  do people behave close to walls?  To choose what is the best strategy
  for moving [e.g.  cooperation (grouping, buddying, etc.) {\em versus}
  selfishness (walking away from groups)] one may also wish to explore
  basic aspects of the dynamics of non-momentum conserving inelastic
  collisions. Billiard dynamics, or biased billiards like those modeling
  the prisoner's dilemma, or broader contexts involving stochastic game
  theory (see \cite{Szilagyi}), perhaps involving non-standard (strongly
  non-Gaussian) scenarios, where energy can be exchanged between
  particles in a non-standard way need to be studied
  \cite{Eggels}. Recall that the Newtonian principle of action and
  reaction is not necessarily true anymore in this framework;
  see \cite{Haret}.

\item[(iii)] A quite similar pile-up effect to the one seen in Figure
  \ref{SSmass} appears as a result of the motion of edge dislocations on
  slip planes in steel plasticity. The dislocations are repulsively
  interacting defects naturally arising in the crystalline structure of
  materials (here dual phase steels). Their motion is typically
  accelerated by the action of a macroscopic stress. As result of this,
  the dislocations are pushed towards a piling-up in the boundary later
  present at the interface between the strong and weak material phase;
  see \cite{dislocationsLucia, dislocationsPatrick} for mathematical
  evidence on the formation of the pile-up starting off from a suitably
  interacting particle system.  Is there a hidden threshold mechanism
  responsible for the formation of the pile-up of dislocations? We
  suspect that the high contrast between the stiffnesses of the two
  steel phases is the responsible threshold. We plan to use a rigorous
  upscaling/homogenization procedure to shed more light on connecting
  density thresholds (high-contrast) with pilling-ups.

\item[(iv)] To which extent cooperation is profitable?  is a basic
  question studied recently for instance in \cite{PLOSPetre}.% by group
  psychologists and socio- econo- physicists. %Essentially, when
  neglecting the effect of population size, thresholds and boundary
  conditions, The authors of \cite{PLOSPetre} are pointing out the
  superiority of collaborative interaction rules as compared to
  follow-the-leader type of interactions, making clear connections
  between concepts like group rationality and deliberative democracy.
  From yet a different perspective, this subject is intimately connected
  to the dynamics of opinions (cf. e.g. the work by S. Galam; to get a
  hint on this see \cite{Galam1,Galam2} and references cited therein) as
  indicated also in \cite{Moshman} (in the spirit that deliberative
  democracy outreasons enlightened dictatorship). One could stretch more
  this idea towards eventual links to percolation theory applied this
  time not to a porous media setting, but rather to dynamically evolving
  networks (societies). We refer the reader to \cite{Rutger}, for some
  preliminary thoughts around the idea of percolation thresholds
  occurring in structured social systems.

  % Resilience in networks? percolation in regenerating tissues,
  % societies? Idea of thrershold in systems \cite{Rutger},

  % modeling panic

  % more on defning the viscosity
\item[(v)] Both the lattice system and the population balances
  approach \`a la Becker-D\"oring share many similarities. However, there are a few essential differences
  between the two approaches. An important one is the following: For
  small $N$, the presence of the threshold $T$ seems to be beneficial
  for the particles leaving the lattice system; however this effect is
  lost completely in the Becker-D\"oring approach (compare Figure
  \ref{Cluster_exit}).  This seems to be due to the choice of boundary
  conditions in the continuum system. On the other hand, we conjecture
  that the continuum limit of the lattice system is a sort of non-linear
  diffusion equation with inherited threshold, while we see that the
  Becker-D\"oring system is not emphasizing the threshold effects when
  changing the size (or nonlinearity) of the effective diffusion
  coefficient (see e.g. Figure \ref{fig:diffusion}). The challenging
  question is here: Derive (and then prove rigorously) the mean-field
  limit for the lattice system. Alternatively, one can reformulate the
  lattice model in terms of myopic random walkers in an exclusion
  process in the spirit of \cite{Landman} and then prove rigorously the
  validity of the corresponding mean-field model (a porous media-like
  equation).

\item[(vi)] Based on our working experience with continuum models with
  distributed microstructures, we expect that it is possible to couple
  the two models for groups dynamic within a single multiscale
  framework. The challenge here is to establish the right micro-macro
  transmission condition (in this case, a discrete-to-continuum
  coupling). We believe that steps in this direction are possible,
  inspired for instance by the way the human language is treated in
  \cite{language} as a hybrid system.

\end{enumerate}
\CCLsection*{Acknowledgments}
We thank Anne Eggels, Joep Evers, Francesca Nardi and Rutger van
Santen (Eindhoven), Petre Cur\c seu (Tilburg) as well as Errico
Presutti (Rome) for fruitful discussions on this and closely related
topics. A.M. thanks the NWO's Complexity program (project "Correlating
fluctuations across the scales"), O.K.  acknowledges financial support
from the EU ITN FIRST project.

\appendix
\CCLsection{Becker-D\"oring system in the context of a two-scale modeling approach}\label{BB}
% {\bf Not ready}

\CCLsubsection{Background}

This section contains a brief derivation of a structured-population
model, which is a special case of a multi-feature continuity equation
cf. \cite{MB}. It provides \ a general framework for some of the
equations we are dealing with. For a related derivation using
densities, see \cite{Perthame}, e.g. At a more general scale the
following considerations yield some sort of a transport equation or
continuity equation, respectively, with two features being involved in
the transport (also: cf. \cite{smol,Odo} et al.). In the present
situation, the "location in the corridor" and the "group size"
constitute the two "features". The first is a continuous, the second a
discrete variable. Our aim is to derive a population-balance equation,
(\ref{A51}), able to describe the evolution of pedestrian groups in
obscured regions.

Fix $N\in%
% TCIMACRO{\U{2115} }%
% BeginExpansion
\mathbb{N}
% EndExpansion
$, let $\Omega\subset%
% TCIMACRO{\U{211d} }%
% BeginExpansion
\mathbb{R}
% EndExpansion
^{2}$ be the dark corridor (open, bounded with Lipschitz boundary), $S$ - the
observation time interval and $K_{d}:=\left\{  0,1,2,3,..,N\right\}  $ - the
collection of all admissible group sizes. We say that a $Y$ belongs to
$K^{\prime}\subseteq K,$ if it belongs is part of some group with a size \ K.
Furthermore, $\mathfrak{A}_{\Omega}:=\mathfrak{B}^{2}(\Omega)$, $\mathfrak{A}%
_{S}:=\mathfrak{B}^{1}(S)$ are the corresponding Borel $\sigma$-algebras with
the corresponding Lebesgue-Borel measures $\lambda_{x}:=\lambda^{2}$ and
$\lambda_{t}:=\lambda^{1},$ respectively; $\mathfrak{A}_{K_{d}}:=\mathfrak{p}(K_{d})$
is equipped with the counting measure $\lambda_{c}^{\prime}(K):=\left\vert
  K^{\prime}\right\vert $. We call $\lambda_{tx}:=\lambda_{t}\otimes\lambda_{x}$
the \textit{space-time measure} and set $\lambda_{txc}:=\lambda_{t}%
\otimes\lambda_{x}\otimes\lambda_{c_{j}},$ $\mathfrak{A}_{\Omega K_{d}%
}:=\mathfrak{A}_{\Omega}\otimes\mathfrak{A}_{K_{d}},$ $\mathfrak{A}_{S\Omega
  K_{d}}:=\mathfrak{A}_{S}\otimes\mathfrak{A}_{\Omega}\otimes\mathfrak{A}%
_{K_{d}}.$

\CCLsubsection{Derivation of the model}

Fix $t\in S,$ let $\Omega^{\prime}\in\mathfrak{A}_{\Omega},$ $K^{\prime}%
\in\mathfrak{A}_{K_{d}},$ $S^{\prime}\in\mathfrak{A}_{S}$, introduce%
\begin{equation}%
  \begin{array}
    [c]{c}%
    \mu_{Y}(t,\Omega^{\prime}\times K^{\prime}):=\text{ number of }Y^{\prime
    }s\text{ present in }\Omega^{\prime}\text{ }\\
    \text{at time }t\text{ and belonging to the group }K^{\prime}\\
  \end{array}
  \label{A10}%
\end{equation}
and two \textit{production quantities}
\begin{equation}%
  \begin{array}
    [c]{c}%
    \mu_{PY\pm}(S^{\prime}\times\Omega^{\prime}\times K^{\prime}):=\text{ number
      of }Y^{\prime}s\text{ which are added }\\
    \text{to (subtracted from) }\Omega^{\prime}\times K^{\prime}\text{ during
    }S^{\prime}\text{ and}\\
    \mu_{PY}=\mu_{PY+}-\mu_{PY-}.
  \end{array}
  \label{A15}%
\end{equation}
Note that these numbers might be non-integer.\ \ \ \ \

Given the nature of the problems we are dealing with, we postulate - as a part
of the modeling-

\begin{description}
\item (P1) \ \ For all $K^{\prime}\in$ $\mathfrak{A}_{K_{d}},$ $\Omega
  ^{\prime}\in$ $\mathfrak{A}_{\Omega}$ $\ $and $t\in S:\mu_{Y}(t,\cdot\times
  K^{\prime})$ and $\mu_{Y}(t,\Omega^{\prime}\times\cdot)$ are measures on their
  respective $\sigma$-algebras $\mathfrak{A}_{\Omega}$ and $\mathfrak{A}_{K_{d}%
  },$ respectively.

\item (P2) $\ \ \mu_{PY\pm}(S^{\prime}\times\Omega^{\prime}\times\cdot),$
  $\mu_{PY\pm}(S^{\prime}\times\cdot\times K^{\prime})$ and $\mu_{PY\pm}%
  (\cdot\times\Omega^{\prime}\times K^{\prime})$ are measures on their
  respective $\sigma$-algebras.
\end{description}

Now, we are in the position to formulate a%
\begin{equation}%
  \begin{array}
    [c]{c}%
    \text{Balance principle:}\\
    \mu_{Y}(t+h,\Omega^{\prime}\times K^{\prime})-\mu_{Y}(t,\Omega^{\prime}\times
    K^{\prime})=\mu_{PY}(S^{\prime}\times\Omega^{\prime}\times K^{\prime})\\
    \text{for all }t,t+h\in S,\text{ }\Omega^{\prime}\times K^{\prime}%
    \in\mathfrak{A}_{\Omega}\times\mathfrak{A}_{K_{d}},\text{ }S^{\prime
    }:=(t,t+h].\\
  \end{array}
  \label{A20}%
\end{equation}
Addition to $\Omega^{\prime}\times K^{\prime}$, modeled by $\mu_{PY+},$ can
happen by addition \textit{inside} of $\Omega^{\prime}\times K^{\prime}$ as
well as by fluxes \textit{into} $\Omega^{\prime}\times K^{\prime}.$ A similar
remark applies to subtraction and $\mu_{PY-}$. This gives rise to assume
$\mu_{PY+}$ to be the sum of an interior production part, $\mu_{PY+}^{int},$
and a flux part, $\mu_{PY+}^{flux}.$ We proceed similarly with $\mu_{PY-}$ and
have, with the
\[
\text{net productions }\mu_{PY}^{int}:=\mu_{PY+}^{int}-\mu_{PY-}^{int}\text{
  and \ }\mu_{PY}^{flux}:=\mu_{PY+}^{flux}-\mu_{PY-}^{flux}:
\]%
\begin{equation}
  \ \ \mu_{PY}=\mu_{PY}^{int}+\mu_{PY}^{flux}=\left(  \mu_{PY+}^{int}-\mu
    _{PY-}^{int}\right)  +\left(  \mu_{PY+}^{flux}-\mu_{PY-}^{flux}\right)  .
  \label{A25}%
\end{equation}

We extend $\mu_{Y}(t,\cdot\times\cdot)$ and $\mu_{PY\pm}(\cdot\times
\cdot\times\cdot)$ by the usual procedure to measures $\overline{\mu}_{Y}%
=\mu_{Y}(t,\cdot)$ and $\overline{\mu}_{PY\pm}=\overline{\mu}_{PY\pm}(\cdot)$
on the product algebras $\mathfrak{A}_{\Omega}\otimes\mathfrak{A}_{K_{d}}$ and
$\mathfrak{A}_{S}\otimes\mathfrak{A}_{\Omega}\otimes\mathfrak{A}_{K_{d}},$ respectively.

Note that the quantities in (P1) and (P2) and the extensions are finite.

The following postulate prevents accumulation on sets of measure zero. It
reads as

\begin{description}
\item (P3) \ $\overline{\mu}_{Y}(t,\cdot)\ll\lambda_{xc}$ (absolutely continuous).
\end{description}

Therefore, for all $t\in S$ there are integrable Radon-Nikodym densities
$u(t,\cdot)=\frac{d\overline{\mu}_{Y}(t,\cdot)}{d\lambda_{xc}}\footnote{Note
  with respect to Section \ref{BDa}: $u_{i}(t,x)$ from Section \ref{BDa} corresponds to
  $u(t,x,i)$ here.},$ i.e.%
\begin{equation}
  \overline{\mu}_{Y}(t,Q^{\prime})=\int_{Q^{\prime}}u(t,(x,i))d\lambda
  _{xc}\text{ \ for all }Q^{\prime}\in\mathfrak{A}_{\Omega K}. \label{A27}%
\end{equation}

The absolute-continuity assumption

\begin{center}
  \begin{figure}[h]
    \centering
    % \begin{picture}(500,180)(-120,0)
    \includegraphics[width=6cm]{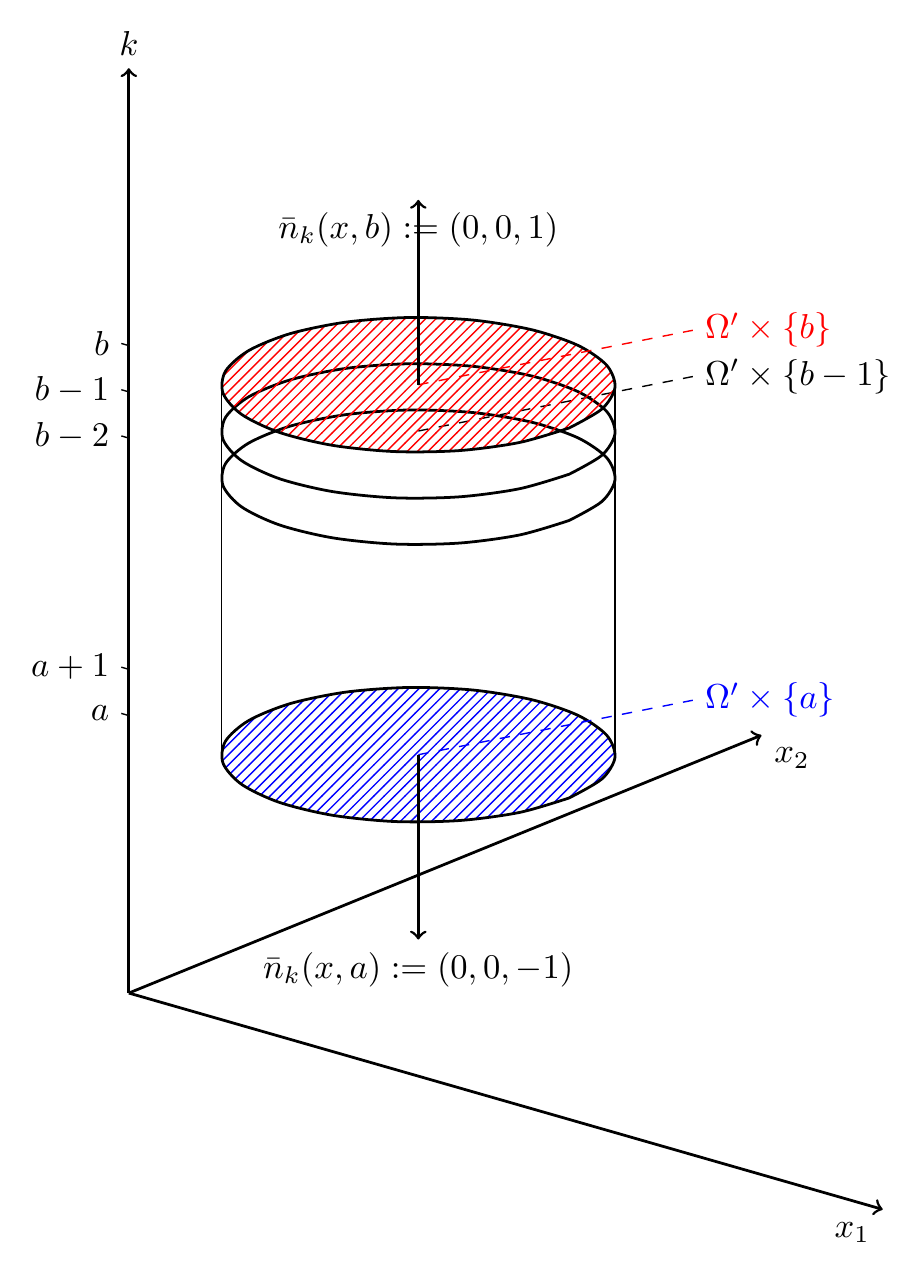}
    % \end{picture}
    \caption{Special interactions regions on the surface $\mathfrak{F}$.}
    \label{FigA1}
  \end{figure}
\end{center}

\begin{description}
\item (P4) \ $\overline{\mu}_{PY}^{int}\ll\lambda_{txc}$
\end{description}

excludes the presence of $Y^{\prime}s$ on sets of $\lambda_{txc_{1}}$-measure
zero. Moreover it assures the existence of the Radon-Nikodym density
\begin{equation}
  f_{PY}^{int}:=\frac{d\overline{\mu}_{PY}}{d\lambda_{txc}}\in L_{loc}%
  ^{1}(S\times\Omega\times K_{d},\mathfrak{A}_{S\Omega K_{d}},\lambda_{txc}).
  \label{A30}%
\end{equation}
In order to get a reasonable idea for a representation of the flux measure we
consider the special case $Q^{\prime}=\Omega^{\prime}\times K^{\prime}$ with,
say, $K^{\prime}=\left\{  a,a+1,...,b\right\}  \in\mathfrak{p}(K_{d}).$ The
"surface"
\[
\mathfrak{F}:=\Omega^{\prime}\times\left\{  a\right\}  \cup\Omega^{\prime
}\times\left\{  b\right\}  \cup\partial\Omega^{\prime}\times K^{\prime}%
\]
is the location of any interaction with the outside of $Q^{\prime}$. There are
two locations on $\mathfrak{F}$ to enter or leave $Q^{\prime}$ from the
outside - one via $\mathfrak{F}_{1}:=\Omega^{\prime}\times\left\{  a\right\}
\cup\Omega^{\prime}\times\left\{  b\right\}  ,$ the other one through
$\mathfrak{F}_{2}:=\partial\Omega^{\prime}\times K^{\prime}$ \ (see Figure \ref{FigA1}).%

The unit-outward normal field $\mathbf{n}=\mathbf{n}(x,\kappa)$ on
$\mathfrak{F}$ can be split into two orthogonal components, $\mathbf{n}%
=\mathbf{n}_{x}+\mathbf{n}_{\kappa},$ $\mathbf{n}_{x}=(n_{x},0),$
$\mathbf{n}_{\kappa}=(0,n_{\kappa}),$ respectively. It is $n_{\kappa}(x,a)=-1$,
$n_{\kappa}(x,b)=+1$ and $n_{x}=n_{x}(x,\kappa)$ is the a.e. existing outward
normal on $\partial\Omega.$ 
Borrowing from the theory of Cauchy interactions, cf. \cite{Schuricht}, e.g..

\begin{description}
\item (P5) we assume\ for all $t\in S$ the existence of two vector fields
  \begin{eqnarray}
    j_{x}(t,\cdot)&:&\Omega\times K_{d}\rightarrow
    \mathbb{R}^2\nonumber\\
    % EndExpansion
    j_{\kappa}&:&\Omega\times K_{d}\rightarrow \mathbb{R}\nonumber
  \end{eqnarray}
  with%
  \[
  \overline{\mu}_{PY}^{flux}:=\overline{\mu}_{PYx}^{flux}+\overline{\mu
  }_{PY\kappa}^{flux}\text{,}%
  \]
  where
  \begin{equation}%
    \begin{array}
      [c]{rl}%
      \text{ }\\
      \overline{\mu}_{PYx}^{flux}(S^{\prime}\times\Omega^{\prime}\times K^{\prime
      })&:=\int_{S^{\prime}}\int_{\mathfrak{F}_{_{2}}}-j_{x}(\tau,x,i)\cdot
      n_{x}(x,i)d\sigma_{x}d\lambda_{c}d\tau\text{ \ }\\
      &\text{and}\\
      \overline{\mu}_{PY\kappa}^{flux}(S^{\prime}\times\Omega^{\prime}\times
      K^{\prime})
      &:=\int_{S^{\prime}}\int_{\Omega^{\prime}}-j_{\kappa}(\tau,x,b)n_{\kappa
      }(x,b)\\&-j_{\kappa}(\tau,x,a)n_{\kappa}(x,a)dxd\tau.\\
    \end{array}
    \label{A35}%
  \end{equation}

\end{description}

In (\ref{A35}),  $\sigma_{x}$ - is the $1D$-surface (= curve length-) measure. $\overline{\mu
}_{PYx}^{flux}(S^{\prime}\times\Omega^{\prime}\times K^{\prime})$ calculates
the net gain/loss of the $Y^{\prime}s$ in $\Omega^{\prime}$ belonging to one
of the size groups from $K^{\prime}$ due to physical motion from/to the
outside of $\Omega^{\prime}$ into/out of $\Omega^{\prime}$.

Furthermore, $\overline{\mu}_{PY\kappa}^{flux}(S^{\prime}\times\Omega^{\prime}\times\left\{  i\right\}
)$ calculates the net gain/loss of the $Y^{\prime}s$ in $\Omega^{\prime}$
belonging to the size group labelled by $i$ due to reasons within $K$. Since,
in the given situation of Section \ref{BDa}, there is no interaction with groups of
size $\kappa>N$ or $\kappa<0$ (these group sizes are not admissible!), we have
to require%
\begin{equation}
  j_{\kappa}(t,x,0)=j_{\kappa}(t,x,N)=0\text{ \ for all }t\in S,\text{ }%
  x\in\Omega.\label{A37}%
\end{equation}

Introducing the \textit{discrete partial derivative} by
\[
\text{ }\partial_{i}^{d}j_{\kappa}(t,x,i):=j_{\kappa}(t,x,i+1)-j_{\kappa
}(t,x,i),\text{ }i\in K
\]
and assuming $u,$ $f_{PY}^{int},$ $\operatorname{div}_{x}j_{x}$ and
$\partial_{\kappa}^{d}j_{\kappa}$ to be sufficiently regular, we obtain%

\begin{align*}
  \overline{\mu}_{PY}^{flux}(S^{\prime}\times Q^{\prime}) &  =\int_{S^{\prime}%
  }\int_{Q^{\prime}}-\operatorname{div}_{x}(j_{x}(\tau,x,i)d\lambda_{c}dxd\tau\\
  &  +\int_{S^{\prime}}\int_{Q^{\prime}}-\partial_{i}^{d}j_{\kappa}(t,x,i)d\tau
  dxd\lambda_{c}.
\end{align*}
Combining (\ref{A20}) - (\ref{A35}), Fubini's theorem, and division by $h$, imply%
\[%
\begin{array}
  [c]{c}%
  \int_{Q^{\prime}}\frac{1}{h}\left(  u(t+h,x,i)-u(t,x,i)\right)  dxd\kappa\\
  =\int_{Q^{\prime}}\frac{1}{h}\int_{t}^{t+h}\left(  f_{PY}^{int}(\tau
    ,x,i)-\left(  \operatorname{div}_{x}j_{x}(\tau,x,i)+\partial_{i}^{d}j_{\kappa
      }(t,x,i)\right)  \right)  d\tau dxd\lambda_{c}.\\
\end{array}
\]

Under appropriate smoothness conditions on $u,$ $f_{PY}^{int},$ $j_{x}$ and
$j_{\kappa\text{ \ }}$we obtain in the limit $h\rightarrow0$ (the classical
continuity equation with a slightly different interpretation of the entries)%

\begin{equation}
  \frac{\partial u}{\partial t}(t,x,i)+\left(  \operatorname{div}_{x}%
    j_{x}+\partial_{i}^{d}j_{\kappa}(t,x,i)\right)  =f_{PY}^{int}(t,x,i).
  \label{A51}%
\end{equation}

% \subparagraph
\CCLsubsection{Connection with the model in Section \ref{BDa}:}

In order to obtain a workable model, one has to specify the flux vectors
$j_{x}$ and $j_{\kappa}$ as well as $f_{PY}^{int}.$ In Section \ref{BDa} this has been
done in (\ref{eq1}) to (\ref{eq6}) by setting

$i=1,...,N$ (there) = $i=1,...,N$ (here), $u_{i}(t,x)$ (there) $=$ $u(t,x,i)$
(here), $-D_{i}\nabla_{x}u_{i}(t,x)$ (there) =$\ j_{x}(t,x,i)\ $(here),\\
$$f_{PY}^{int}(t,x,i) \mbox{(here)} =
\left\{ \begin{array}{lr}
    \sum\nolimits_{i=1}^{N}
    \alpha_{i}u_{i}-\sum\nolimits_{i=1}^{N}
    +\beta_{i}u_{i}u_{1} & \mbox{ if }   i=1,\\
    \beta_{i-1}u_{i-1}u_{1}-\beta_{i}u_{i}u_{1} & \mbox{ if }  i\in\{2,...,N-1\},\\
    \beta_{N}u_{N-1}u_{1} & \mbox{ if }  i=N,\\
  \end{array}
\right.
$$respectively.

The discrete
derivative $j_{\kappa}(t,x,i)$ (here) corresponds to
\begin{eqnarray}
  j_{\kappa
  }(t,x,i)& =& -\alpha_{i}u_{i}(t,x), \ i=1,2,...,N-1,\nonumber\\
  j_{\kappa}(t,x,0)& =& j_{\kappa}(t,x,N)=0.\nonumber
\end{eqnarray}

% \subparagraph
\CCLsubsection{Derivation of the model in Section \ref{BDa}:}

Specifying $j_{x}(t,x,i)$ as some sort of a diffusion flux in the manner above
means: Individual groups of size $i$ recognize whether a group of the same
size is in their immediate neighborhood and they tend to avoid moving into
the direction of such groups. Employing a Fickian law seems to be the simplest
way to model this.   $f_{PY}^{int}$ models interactions (= merging) \ between
groups of size $i\in K$ and "groups" of size $i=1$: If a single (i.e. a group
of size one) hits a group of size $i<N$, then it might happen, that this
single merges with the group. This turns the group into a group of size $i+1$
and leads to a "gain" for groups of size $i+1$ (modeled by $+\beta_{i}%
u_{i}u_{1}$) and a loss for groups of size $i$ (modeled by $-\beta_{i}%
u_{i}u_{1}$). In any such joining situation the group with $i=1$ looses members
(modeled by $-%
% TCIMACRO{\tsum \nolimits_{i=2}^{N}}%
% BeginExpansion
{\textstyle\sum\nolimits_{i=2}^{N}}
% EndExpansion
\beta_{i}u_{i}u_{1}$). Note, that this model allows only for direct
interaction between groups of size $i$ with groups of size $1$! The $\alpha
$-terms model some "degradation" effect: It might happen, that an individual
leaves a group of size $i\geq2$. This leads to a loss for the groups of size
$i$ (modeled by $-\alpha_{i}u_{i}$), a gain for the groups of size $i-1$ and
a gain for the groups with $i=1$ (modeled by $%
% TCIMACRO{\tsum \nolimits_{i=2}^{N}}%
% BeginExpansion
{\textstyle\sum\nolimits_{i=2}^{N}}
% EndExpansion
\alpha_{i}u_{i}$). $\alpha_{i},$ $\beta_{i}\geq0$ and $D_{i}>0$ are empirical
and assumed to be constant.

Note: In the abstract approach the degradation terms express a flux rather
than a volume source or sink. In the same way as aging can be seen as a flux
("people change their age group by aging with (speed 1)" ) $Y$'s change their
size group by "degradation" of their group. Nevertheless: For fixed $i$,  the
expressions $\alpha_{i}u_{i}$ and $\alpha_{i-1}u_{i-1}$ still remain "volume
sources" and "sinks", respectively. It's just two different ways to look at the same thing.

\bigbreak\noindent
% The bibliography generated by the \textsf{plainnat}   style looks like the following one.

\bibliographystyle{plainnat}
\bibliography{smoke_book1}
\end{document}